\pdfoutput=1
\documentclass[10pt,conference, numbers]{IEEEtran}

\usepackage{xr}

\usepackage{todonotes}
\usepackage{graphicx}
\usepackage{microtype}
\usepackage{hyperref}

\usepackage{mathtools}
\usepackage{amsmath}
\usepackage{amssymb}
\usepackage{amsthm}
\usepackage[altpo]{backnaur}
\usepackage{listings}

\usepackage{multirow}
\usepackage{tabularx,adjustbox,booktabs}
\usepackage{siunitx}
\usepackage{threeparttable}

\usepackage{dcolumn}

\usepackage{array}

\usepackage[]{xcolor}
\usepackage{colortbl}

\usepackage[para]{footmisc}

\usepackage{xcolor}

\usepackage[justification=centering]{caption}

\usepackage[breakable, skins,xparse]{tcolorbox}

\usepackage{tikz}

\usepackage{pifont}

\usepackage{natbib}

\usepackage{listings}

\usepackage{rotating}

\usepackage[ruled,vlined,linesnumbered,inoutnumbered,noend]{algorithm2e}

\usepackage{tikz-cd}

\hypersetup{
    colorlinks = true,
    linkcolor = blue,
    filecolor = blue,      
    urlcolor = blue,
   citecolor = blue,
}

\newcolumntype{$}{>{\global\let\currentrowstyle\relax}}
\newcolumntype{^}{>{\currentrowstyle}}

\newcommand{\suppweb}{\hyperref[app:supplementary]{appendix}}

\newcommand{\suppwebrepo}{\href{https://github.com/se-sic/icse_model_completion}{supplementary website}}

\usepackage[english]{babel}

\makeatletter
\renewcommand\@makefnmark{\hbox{\@textsuperscript{\normalfont\color{blue}\@thefnmark}}}
\makeatother

\usetikzlibrary{shapes.geometric, shapes,arrows, graphs, matrix}

\RequirePackage{titlesec}
\titlespacing*{\section}{2pt}{1.5ex}{1.5ex}
\titlespacing*{\subsection}{2pt}{1ex}{1ex}

\newcolumntype{L}{>{\phantom{$-$}}l}       

\newcommand\restr[2]{{
  \left.\kern-\nulldelimiterspace 
  #1 
  \vphantom{\big|} 
  \right|_{#2} 
  }}

\newcommand{\mde}{model-driven engineering}
\newcommand{\mmodel}{metamodel}

\newcommand{\eo}{edit operation}
\newcommand{\Eo}{Edit operation}

\newcommand{\scg}{simple change graph}
\newcommand{\Scg}{Simple Change Graph}

\newcommand{\nbExperiments}{four}
\newcommand{\trainratio}{75\%}
\newcommand{\testratio}{25\%}

\newcommand{\numbersamplessynthetic}{210}
\newcommand{\numbersamplesrevision}{221}
\newcommand{\numbersamplesindustry}{200}

\newcommand{\numbersamplesforbaselinecomparison}{51}

\newcommand{\ragLong}{retrieval-augmented generation}

\newcommand{\rag}{\textsc{RaMc}}

\newcommand{\SCG}{SCG}

\newcommand{\ceo}{completion operation}
\newcommand{\edgel}{EdgeList}

\newcommand{\generationPhase}{generation phase}
\newcommand{\trainingPhase}{training phase}

\newcommand{\GenerationPhase}{Generation Phase}
\newcommand{\TrainingPhase}{Training Phase}

\newcommand{\ada}{{\small \sf text-ada-001}}
\newcommand{\curie}{{\small \sf text-curie-001}}
\newcommand{\davinci}{{\small \sf text-davinci-003}}
\newcommand{\datasetadacompletion}{\textsc{Batch 1}}
\newcommand{\datasetcuriecompletion}{\textsc{Batch 2}}

\newcommand{\averageTokenAccuracy}{96.9\%}
\newcommand{\minTokenAccuracy}{92.1\%}
\newcommand{\maxTokenAccuracy}{99.0\%}

\newcommand{\semanticCorrectIndustryPercent}{62.30\%}
\newcommand{\typeCorrectSyntheticPercent}{86.19\%}

\newcommand{\totalCost}{347\$}

\newcommand{\numbersamples}{122--221}
\newcommand{\numberindustrysamples}{122}
\newcommand{\maxfewshot}{12}

\newcommand{\figref}[1]{Figure~\ref{#1}}
\newcommand{\secref}[1]{Section~\ref{#1}}
\newcommand{\defref}[1]{Definition~\ref{#1}}
\newcommand{\tblref}[1]{Table~\ref{#1}}

\newcommand{\generalRQ}{RQ 1}
\newcommand{\semanticretrievalRQ}{RQ 2}
\newcommand{\baselineRQ}{RQ 3}
\newcommand{\limitationsRQ}{RQ 4}
\newcommand{\finetuningRQ}{RQ 5}

\newcommand{\generalExp}{Experiment 1 }
\newcommand{\generalExpwithout}{Experiment 1}
\newcommand{\semanticretrievalExp}{Experiment 2 }
\newcommand{\baselineExp}{Experiment 3 }
\newcommand{\limitationsExp}{Experiment 4 }
\newcommand{\finetuningExp}{Experiment 5 }

\newcommand*\circled[1]{\ding{\inteval{#1 + 171}}}
\newcommand{\circleda}{\textcircled{a}}
\newcommand{\circledb}{\textcircled{b}}

\usepackage{ifthen}

\newboolean{changed}
\setboolean{changed}{true}
\newcommand{\changed}[2]{\ifthenelse{\boolean{changed}}
{\color{black}#1\color{black}}{\color{black}#2\color{black}}}

\newboolean{changedFinalRevision}
\setboolean{changedFinalRevision}{true}
\newcommand{\changedFinalRevision}[2]{\ifthenelse{\boolean{changedFinalRevision}}
{\color{black}#1\color{black}}{\color{black}#2\color{black}}}

\newboolean{changedUnderConstruction}
\setboolean{changedUnderConstruction}{true}
\newcommand{\changedUnderConstruction}[2]{\ifthenelse{\boolean{changedUnderConstruction}}%
{\color{black}#1\color{black}}{\color{black}#2\color{black}}}

\newcommand{\pluseq}{\mathrel{+}=}

\newboolean{cut}
\setboolean{cut}{true}
\newcommand{\cut}[2]{\ifthenelse{\boolean{cut}}
{\color{black}#1\color{black}}{\color{black}#2\color{black}}}
{

\definecolor{darkblue}{rgb}{0,0,.75}

\definecolor{eminence}{RGB}{108,48,130}

\lstloadlanguages{Matlab} 
\lstnewenvironment{PseudoCode}[1][]
{\lstset{language=Matlab,basicstyle=\scriptsize, keywordstyle=\color{darkblue},numbers=left,xleftmargin=.04\textwidth,#1}}
{}

\theoremstyle{definition}
\newtheorem{definition}{Definition}[section]
\newtheorem*{remark}{Remark}

\setlength{\tabcolsep}{3pt}

\title{Software Model Evolution with Large Language Models:
Experiments on Simulated, Public, and Industrial Datasets}

\author{
    \IEEEauthorblockN{Christof Tinnes\textsuperscript{\textsection}}
    \IEEEauthorblockA{
        \textit{Siemens AG} \\
        Garching bei München, Germany \\
        christof.tinnes@siemens.com
    }
    \and
    \IEEEauthorblockN{Alisa Welter\textsuperscript{\textsection}}
    \IEEEauthorblockA{
        \textit{Saarland University} \\
        Saarbrücken, Germany \\
        welter@cs.uni-saarland.de
    }
    \and
    \IEEEauthorblockN{Sven Apel}
    \IEEEauthorblockA{
        \textit{Saarland University} \\
        Saarbrücken, Germany \\
        apel@cs.uni-saarland.de
    }
}

\begin{document}
\maketitle              
%

\begingroup\renewcommand\thefootnote{\textsection}
\footnotetext{Equal contribution}
\endgroup

\begin{abstract}


Modeling structure and behavior of software systems plays a crucial role in the industrial practice of software engineering. As with other software engineering artifacts, software models are subject to evolution. Supporting modelers in evolving software models with recommendations for model completions is still an open problem, though.
In this paper, we explore the potential of large language models for this task. In particular, we propose an approach, \rag{}, leveraging large language models, model histories, and \ragLong{} for model completion. Through experiments on three datasets, including an industrial application, one public open-source community dataset, and one controlled collection of simulated model repositories, we evaluate the potential of large language models for model completion with \rag{}.
We found that large language models are indeed a promising technology for supporting software model evolution (\semanticCorrectIndustryPercent{} semantically correct completions on real-world industrial data and up to \typeCorrectSyntheticPercent{} type-correct completions).
The general inference capabilities of large language models are particularly useful when dealing with concepts for which there are few, noisy, or no examples at all.
\end{abstract}

\section{Introduction}\label{sec:introduction}

Models play an important role in modern software and system development~\cite{RodriguesDaSilva2015}, 
software documentation~\cite{kruchten1995, UML2017}, 
system architecture~\cite{SysML2019}, 
simulation~\cite{dabney2004mastering}, 
and industrial automation~\cite{iec61131plc}. 
In practice, all artifacts in software and system development are subject to evolution, which also applies to \emph{software models}\footnote{In our work, to avoid confusion, it's crucial to differentiate between software models and machine learning models.}: 
Software models must evolve because of changing requirements, but they are also subject to bugfixes and refactorings~\cite{visser2007model}. 



From the perspective of a modeling tool, we can understand the evolution of a software model as a sequence of \emph{\eo{s}}:
To change or evolve the model, the user executes \eo{s} (e.g., using mouse clicks and keyboard strokes) provided by the tool. 
Supporting tool users in accomplishing various software model (evolution) tasks is clearly desirable in practice~\cite{deng2016recommendation,Tinnes2021}. 
For the evolution of software models, modeling tools typically provide an initial set of \eo{s} (e.g., adding an attribute to a model element). 
Nevertheless, since the usage of a (domain-specific) language is also subject to evolution and since (project-specific) usage patterns might emerge, this initial set of \eo{s} is likely not exhaustive. 
For example, in object-oriented design, design patterns~\cite{gamma1995design} are widely used and are not part of UML~\cite{UML2017}, but could be provided as \eo{s} by a UML modeling tool. 


For source code, modern integrated development environments already support writing and evolving source code by \emph{(auto-)completion}.
Most notably, the use of large language models (LLMs) has become state-of-the-art for the auto-completion of source code~\cite{chen2021codex,xu2022systematic,ahmad2021unified,feng2020codebert,ahmed2022few,wang2022no}.


The world of software models seems to be lagging behind, and no general approach for software model auto-completion is ready for industrial application.
It has been even argued that the so-called cognification of use cases in model-driven software engineering might turn the difference between (perceived) added value and cost from negative to positive~\cite{cabot2018cognifying}.

\textbf{Problem Statement.}
Notably, for a few domain-specific languages, rule-based approaches exist that use pre-defined \eo{s} or patterns for model completion~\cite{kuschke2017rapmod, kuschke2013recommending, heinemann2012facilitating, sen10completion}.
Using a specification language for defining \eo{s} poses three challenges, though.
First, specifying new \eo{s} requires knowledge about the specification language and the domain-specific language. 
Second, domain-specific \eo{s} are often not explicitly known, that is, they are a form of tacit knowledge~\cite{Polanyi1958}. Externalizing the knowledge is hard or even impossible for domain experts. 
Third, \eo{s} can change over time, for example, because the \mmodel{} changes. 
In the light of these challenges, mining approaches that retrieve \eo{s} are especially appealing, since they do not require any manual specification, no hand-crafting of examples (as in model transformation by example~\cite{Varro2006,Kehrer2017}), and they are not limited to well-formedness rules that can be derived out of the \mmodel{}. Unfortunately, existing approaches such as applying frequent subgraph mining to software model repositories are not scalable~\cite{Tinnes2021}, 
and mining approaches lack abstraction capabilities~\cite{Tinnes2021}. 

Clearly, from the perspective of software model evolution, it is desirable to have \emph{context-dependent auto-completions}, rather than utilizing a fixed set of \eo{s}. 
We posit that generative language models exhibit a deep understanding of language and hold comprehensive knowledge across various domains, which is a result of their training on vast corpora. This capability enhances their potential to interpret and complete software models effectively, which usually encompass a vast amount of natural language data.

While recent research suggests that LLMs could be utilized for model completion~\cite{chaaben2023towards}, we go beyond and utilize model evolution data from model repositories to capture real-world complexities.
It is important to note that, in our work, we explicitly acknowledge the complexity of real-world data, which is due to the close collaboration with our industry partner (who also contributes a case study).

\textbf{Contributions.}
By leveraging existing software model histories\footnote{Note that we use the terms \emph{software model repositories} and \emph{software model histories} interchangeably, and we assume that the repository contains several revisions of a software model.}, and by defining an encoding for serializations of model difference graphs, we study to what extent \ragLong{}, (i.e., we provide examples as context in the prompt) can be used for software model completion.
We find that \rag{} is indeed a promising approach for software model completion, with \semanticCorrectIndustryPercent{} of semantically correct completions. \changedFinalRevision{We furthermore propose to use fine-tuning (i.e., the LLM's weights are adapted by training on parts of our data) for software model completion and compare it to our retrieval-based approach, \rag{}. }{}
LLM's general inference capabilities prove especially helpful in handling noisy and unknown context, and real-time capabilities enabled by LLMs are beneficial for stepwise model completion. 
We conclude that using LLMs for software model completion is viable in practice (despite various complexities), but further research is necessary to provide more task and domain knowledge to the LLM. 

In summary, we make the following contributions:
\begin{itemize}
    \item As a foundation for applying LLMs, we formalize the concept of software model completion based on change graphs and their serialization.
    \item We propose a \ragLong{} approach, \rag{}, for software model completion.
    \item We evaluate \rag{} qualitatively and quantitatively on three datasets, including an industrial application, one public open-source community dataset, and one controlled collection of simulated model repositories. 
    \changedFinalRevision{We compare our approach with the most recent advancements in model completion~\cite{chaaben2023towards} as well as to the alternative of fine-tuning a pre-trained LLM. }{}
    We find that, for all three datasets, LLMs are a promising technology for software model completion, with up to \typeCorrectSyntheticPercent{} correct completions (for the synthetic dataset) and \semanticCorrectIndustryPercent{} of semantically correct completions on the industrial dataset. \changedFinalRevision{Notably, our approach improves significantly over the state of the art~\cite{chaaben2023towards}. }{}
     \changedFinalRevision{Furthermore, it appears that fine-tuning 
     can be an alternative to \ragLong{} that is worthwhile investigating.}{}
\end{itemize}
Source code for the experiments, scripts, public datasets, and results are publicly available (see Section~\ref{sec:data_availability}).

\section{Related Work}
 
Various approaches have been proposed for software model completion, ranging from rule-based approaches to data mining techniques and more sophisticated machine learning approaches. An overview of recommender systems in model-driven engineering is given by Almonte et al.~\cite{almonte22}.
Some of the previous work studies recommending model completions by utilizing knowledge bases such as pattern catalogs or knowledge graphs~\cite{agt2018domore,kuschke2017rapmod, kuschke2013recommending,maeder2021,li2013efficient,deng2016recommendation,minas2009}. 
Consequently, these research efforts are often domain-specific, as they require the provision of domain-specific catalogs (a.k.a., the cold start problem), such as for UML~\cite{kuschke2017rapmod,kuschke2013recommending,maeder2021} or business process modelling~\cite{deng2016recommendation, li2013efficient}.

Another common approach is to use already existing model repositories and employ techniques such as frequency-based mining, association rule mining, information retrieval techniques, and clustering to suggest new items to be included in the model~\cite{adhikari2023simima,stephan2019towards,di2023memorec,elkamel2016uml} or new libraries for use~\cite{heinemann2012facilitating}. 
\changedFinalRevision{MemoRec~\cite{di2023memorec} and MORGAN~\cite{di2023morgan} are frameworks that use a graph-based representation of models and a similarity-based information retrieval mechanism 
to retrieve relevant items (such as classes) from a database of modelling projects. However, their graph-based representation does focus on the relationship between a model element and its attributes, but it does not capture relationships \emph{between} different elements in the model and consequently may not capture the essential semantics and constraints of the model and modelling languages. } 

Repository mining and similarity-based item recommendation techniques are often combined~\cite{deng2016recommendation, li2013efficient}. 
K{\"o}gel et al.~\cite{kogel2016automatic,kogel2017recommender} identify rule applications in current user updates and find similar ones in the model's history.
More generally, one could automatically compute consistency-preserving rules~\cite{Kehrer2016} or pattern mining approaches~\cite{Tinnes2021,tinnes2023mining,langer2013posteriori} to derive a set of rules to be used in conjunction with a similar association rule mining approach. 

Another strategy to generate model completion candidates that comply with the given \mmodel{} and additional constraints involves using search-based techniques~\cite{steimann2013}. 
Without knowledge about higher-level semantics, these approaches are more comparable to the application of a catalog of minimal consistency-preserving \eo{s}~\cite{Kehrer2016}.


Regarding the application of natural language processing (NLP)~\cite{burgueno2021nlp} and language models~\cite{tsigkanos23, weyssow2022recommending},
Burgue{\~n}o et al.~\cite{burgueno2021nlp} propose an NLP-based system using word embedding similarity to recommend domain concepts.
Weyssow et al.~\cite{weyssow2022recommending} use a \changedFinalRevision{transformer-based language model }{} to recommend \mmodel{} concepts without generating full model completions. \changedFinalRevision{
Di Rocco et al. \cite{di2022finding} introduce a recommender system using an encoder-decoder neural network to assist modelers with editing operations. It suggests element types to add, but leaves the specification of details, values, and names of these elements and operations to the human modeler. Gomes et al. \cite{gomes2023dome} use natural language processing to translate user intents, expressed in natural language, into actionable commands for developing and updating a system domain model. }{}
While code completion and model completion are closely related, recent research has mainly concentrated on code completion, where LLMs seem to be the state of the art~\cite{chen2021codex,jesse2023large,ciniselli22,Sobania2021GPvsGPT}. 
Considering the close connection to code and model completion, it's essential for us to explore further how generative approaches, such as LLMs, operate within the context of software model completion of complex real-world models. 
Most closely to this work, is an approach by Chaaben et al.~\cite{chaaben2023towards}, which utilized the few-shot capabilities of GPT-3 for model completion \changedFinalRevision{by providing example concepts of unrelated domains.  }{}
In contrast, our approach takes a different avenue, leveraging model evolution from model repositories.
Cámara et al.~\cite{camara2023assessment} further extend on Chaaben et al.'s research by conducting experiments to assess ChatGPT's capability in model generation. 
Ahmad et al.\ explore the role of ChatGPT in collaborative architecting through a case study focused on defining Architectural Significant Requirements (ASRs) and their translation into UML~\cite{ahmad2023towards}. 
\changedFinalRevision{
In the \suppweb{}, a table summarizing related work on model completion is provided. }{}

A slightly different but similar research area focuses on model repair \cite{neubauer2017automated,iovino2020model,nassar2017rule,mazanek2009generating, ohrndorf2021history, stephan2019towards}. 
\textsc{ReVision}~\cite{ohrndorf2021history} uses so-called consistency-preserving edit operations to detected inconsistencies and then uses the pre-defined \eo{s} to recommend repair operations. 


\section{Formal Definitions}\label{sec:background}
In this section, we describe the fundamental concepts essential for the subsequent approach and analysis.

\subsection{Software Models, Edit Operations and Model Completion}\label{sec:eos}

In \mde{}, the language for a software model (i.e., its abstract syntax and static semantics) is typically defined by a \mmodel{} $\mathit{TM}$.
We denote by $\mathcal{M}$ the set of all valid models (according to some \mmodel{}). 
This can be formalized using typed attributed graphs~\cite{Biermann2012,Ehrig2004}.
\begin{definition}[Abstract Syntax Graph]\label{def:asg}
    An \emph{abstract syntax graph} $G_m$ of a model $m \in \mathcal{M}$ is a attributed graph, typed over an attributed type graph $TG$ given by \mmodel{} $\mathit{TM}$.
\end{definition}

The idea of typed graphs is to define a graph homomorphism (i.e., a function from the typed graph $G$ to the type graph $TG$). Details of this formalization are given by Biermann et al.~\cite{Biermann2012}.
The abstract syntax graph of a model and its type graph contain all information that a model holds. 
In this paper, we are concerned with model repositories. We assume that the modelling tool takes care of checking the correct typing of the software models. 
Furthermore, we work with a simplified graph representation of the models in which the abstract syntax graph is a \emph{labeled directed graph} with node and edge labels equal to a textual representation of corresponding classifiers and relationships of the abstract syntax graph (cf.~\defref{def:asg}). 

\begin{definition}[Labeled Directed Graph]\label{def:labeled_graph}
    Given a label alphabet $L$, a \textit{labeled directed graph} $G$ is a tuple $(V,E,\lambda)$, where $V$ is a finite set of nodes, $E$ is a subset of $V \times V$, called the edge set, and $\lambda: V \cup E \to L$ is the labeling function, which assigns a label to nodes and edges. 
\end{definition}

Rather than working directly on the abstract syntax graph of the models, we will mostly be working with model differences.
\begin{definition}[Structural Model Difference]
    A \emph{structural model difference} $\Delta_{mn}$ of a pair of model versions $m$ and $n$ is obtained by matching corresponding model elements in the model graphs $G_{m}$ and $G_{n}$ (using a model matcher~\cite{stephan2013survey}, e.g., EMFCompare~\cite{brun2008model} or SiDiff~\cite{schmidt2008constructing}). There are added elements (the ones present in $G_{n}$ but not in $G_{m}$), removed element (the ones present in $G_{m}$ but not in $G_{n}$), and preserved elements which are present in $G_{m}$ and $G_{n}$.
\end{definition}

We assume that this matching is deterministic, that is, given two models $m, n \in \mathcal{M}$, we obtain a unique structural model difference $\Delta_{mn}$.
The difference can be represented as a {\em difference graph} $G_{\Delta{mn}}$~\citep{ohrndorf2021history}. More concretely, we add the change type ( ``Add'', ``Preserve'', or ``Remove'') in the node and edge labels, and matching elements (i.e., the preserved ones) from $G_{m}$ and $G_{n}$ are unified (present only once). 

We define a \emph{\scg{}} to be the smallest subgraph comprising all changes in the difference graph $G_{\Delta_{mn}}$. 

\begin{definition}[\Scg{}] \label{definition_scg}
Given a difference graph $G_{\Delta_{mn}}$, a \emph{\scg{}} $SCG_{\Delta_{mn}} \subseteq G_{\Delta_{mn}}$ is derived from $G_{\Delta_{mn}}$ by first selecting all the elements in $G_{\Delta_{mn}}$ representing a change (i.e., added, removed nodes and edges) and, second, adding preserved nodes that are adjacent to a changed edge. 
\end{definition}

\begin{definition}[Endogenous model transformation]
An \emph{endogenous model transformation} is a pair $t = (m,n) \in \mathcal{M} \times \mathcal{M}$. We call $m$ the \emph{source model} and $n$ the \emph{target model} of the transformation and $\mathcal{T} \stackrel{\text{def}}{=}\mathcal{M} \times \mathcal{M}$ the space of endogenous model transformations.
\end{definition}

Next, we define a function $SCG\colon \, \mathcal{T} \to \mathcal{G}$ that takes a model transformation (i.e., a pair of models) as input and returns the simple change graph for the corresponding model difference.
We can use $SCG$ to define an equivalence relation on $\mathcal{T}$ by
\begin{equation*}
t_1 = (m,n) \sim t_2 = (k,l) \,  \iff \,
SCG_{\Delta_{mn}} = SCG_{\Delta_{kl}}.
\end{equation*}
It is straightforward to see that this relation indeed defines an equivalence relation (i.e., the relation is reflexive, symmetric, and transitive). We can therefore define the quotient set $\mathcal{T}/{\sim}$. By construction there is bijection from the quotient set to the range of $SCG$. We can therefore use this construction to formally define the concept of an \emph{\eo{}}.

\begin{definition}
    An \emph{\eo{}} is an equivalence class in the set $\mathcal{E}\stackrel{\text{def}}{=} \mathcal{T}/{\sim}$. An \eo{} is therefore a set of model transformations that have the same simple change graph.
\end{definition}
\begin{figure}[htbp]
	 		\centering
			\includegraphics[width=\columnwidth]{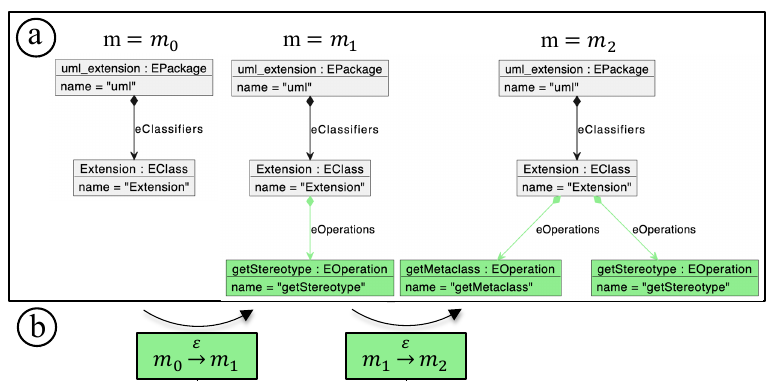}
			\caption{Visual presentation of our  example taken from the \textsc{RepairVision} dataset: \circleda{} Evolutionary View: User performs edit operations one by one. \circledb{} Evolution can be performed by a user or by using a completion approach.}
			\label{fig:motivatingexample}
\end{figure}

\changedFinalRevision{\begin{remark}
    The graph labeling function $\lambda$ allows us do define the scope of the \eo{}. 
    For example, if we are interested only in the type of nodes and edges, we can omit the attributes from the label.
    Likewise, if we are interested in the attributes, or only want to set them during execution time, we can define placeholders for the attribute values in the labels.
    Therefore, we define \eo{s} only up to the concrete label representation, which leaves some freedom for templating. In this work, we do make use of placeholders only during the evaluation (e.g., checking for type correctness).
\end{remark}}

Given an \eo{} $\varepsilon$ and a model $m$, one can perform the removal of ``Remove'' nodes and the gluing of ``Add'' nodes as defined by the \scg{} corresponding to $\varepsilon$, and then set concrete attributes. This yields the corresponding model $n$ with $(m, n) \in \varepsilon$. This way, an \eo{} $\varepsilon \in \mathcal{E}$ can be interpreted as a template for a model transformation, which is in line with previous constructions \cite{Biermann2012,Kehrer2015,Tinnes2021}. We write $m \stackrel{\varepsilon}{\to} n$ to denote a concrete element (i.e., a model transformation) in the equivalence class $\varepsilon \in \mathcal{E}$.
We are interested in completing software models. That is, for an existing evolution $m \stackrel{\varepsilon}{\to} n$, \changedFinalRevision{ we want to find a completion $\gamma \in \mathcal{E}$, such that $m \stackrel{\varepsilon}{\to} n \stackrel{\gamma}{\to} c$ is a realistic completion, meaning, in some real-world scenarios, it actually will be done by a modeler.}{}

\begin{definition}[Model Completion]
    Given a set of model transformations $\mathcal{T}$, \emph{model completion} is a computable function $C\colon \mathcal{T} \to \mathcal{T}$ that, given a model transformation $m \stackrel{\varepsilon}{\to} n$ from a source model $m$ to a (partial) target model $n$, computes a model transformation $C(m \stackrel{\varepsilon}{\to} n) = n \stackrel{\gamma}{\to} c$. We call the \eo{} $\gamma$ 
    a \emph{software model completion}.  
\end{definition}
\changedFinalRevision{

Given a model completion $\gamma$, we denote the application of $\gamma$ to model $n$ by $\pi: \mathcal{M} \times \mathcal{E} \to \mathcal{T}$, where $\pi{(m, \gamma \circ \varepsilon)} = (n,c)$. In general, for an \eo{} $\varepsilon$, there might be zero or more applications to a given model $m \in \mathcal{M}$. Nevertheless, given that the matching in $n$ is fully defined by the application of $\varepsilon$, there is a uniquely defined candidate $(n,c) \in \mathcal{T}$. 
}

\subsection{Language Models}\label{sec:lm}

Language models, as \emph{generative models}, have the capability to produce new sequences of text based on their training data.
 
\begin{definition}[Language Model]
    A \emph{language model} is a conditional probability distribution $\mathbb{P}(\omega|c)$ for a (sequence of) token(s) $\omega$, given a sequence of context tokens $c$. 
\end{definition}

\vskip 1ex
The probability distribution is typically derived from a \emph{corpus} of documents, containing (some of) the tokens.
With the success of transformer architecture \cite{vaswani2017attention}, LLMs have become quite popular now and are used in plenty of domains including software engineering~\cite{samoaa2022systematic,zhao2021natural,Xu2022LLMCode}.
There are two tactics available to feed domain knowledge or context into a generative language model: fine-tuning and \ragLong{}.
Retrieval-augmented generation includes additional knowledge in the context (or prompt). Fine-tuning adjusts the LLM's weights based on additional training data. 



\section{Approach} \label{sec:approach}
In this section, we describe \rag{} -- our approach of \emph{how} to employ LLMs to (auto-)complete software models.

\subsection{Running Example }\label{sec:motivation}

Consider the motivating example depicted in \figref{fig:motivatingexample}, 
which originates from one of our datasets, \textsc{RepairVision}, further explained in \secref{sec:dataset}. 
In \circleda{}, we show the evolution of its abstract syntax graph\footnote{Due to obvious space constraints, only a small part of the original model (only one out of 256 classifiers and 2 out of 741 operations) is shown}. In this evolution scenario, a modeller adds the UML Profiles mechanism (cf. UML specification~\cite{UML2017}, Chapter 12.3) to the \textsc{Ecore} \mmodel{}\footnote{UML, according to the Meta-Object Facility \cite{omg2013mof}, is itself a model according to its meta-metamodel, \textsc{Ecore}, and therefore covered by the present work.} of UML 2.5.1. 
Step by step the modeller extends the existing UML \mmodel{} with additional functionality, currently focusing on the {\small \sffamily EClass} extension in the {\small \sffamily UML} package.
In a first step, the modeller adds an operation {\small \sffamily getStereotype} (responsible for accessing the  {\small \sffamily Sterotype} of the extensions associated with an element in the (meta-)model). As defined in the UML specification~\cite{UML2017}, every {extension} has access to the {\small \sffamily Metaclass} it extends, realized in \textsc{Ecore} by the {\small \sffamily EOperation} {\small \sffamily getMetaclass}.
This {\small \sffamily EOperation} is implemented by the modeller in a second step.
These steps in the evolution of the UML \mmodel{} could be performed via \eo{s} by a human user, or likewise, recommended in the form of a model completion (as depicted in \circledb{} of \figref{fig:motivatingexample}).

\begin{figure}[htbp]
	 		\centering
			\includegraphics[width=\columnwidth]{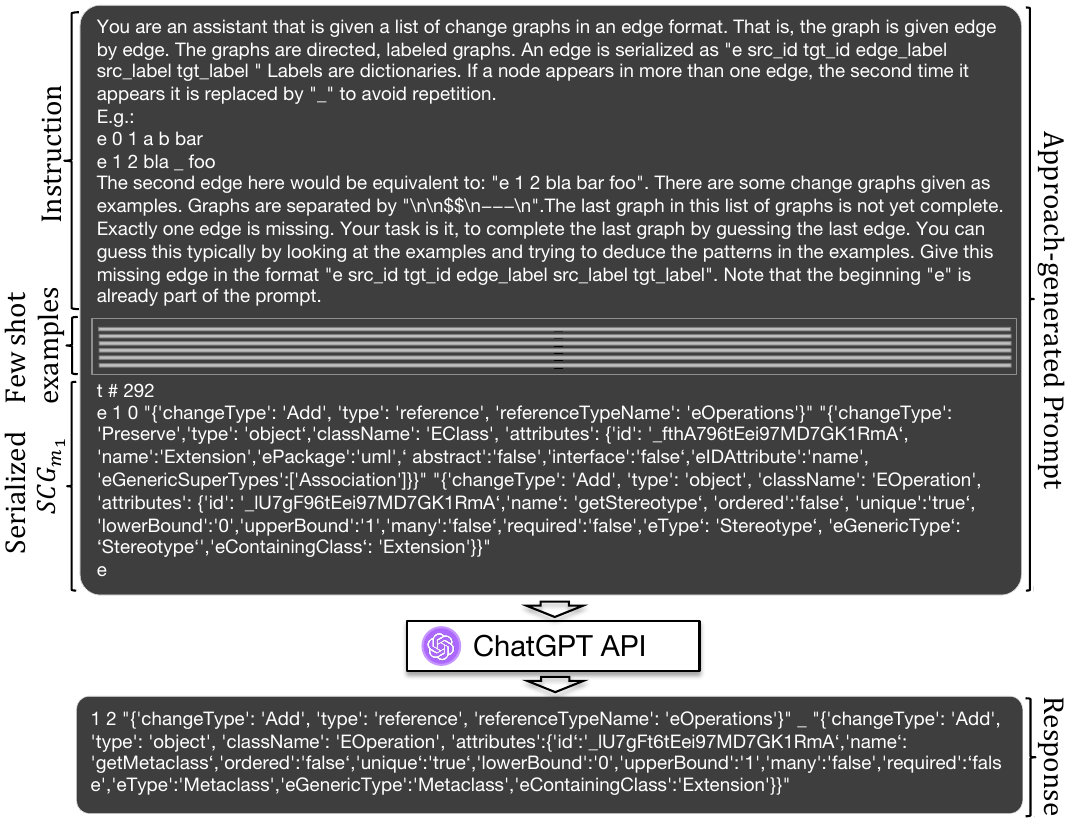}

			\caption{Detailed prompt and \scg{} serialization of the \rag{} approach corresponding to the example given in \figref{fig:motivatingexample}, exact few-shot examples are provided in the \suppweb{}.}
			\label{fig:serializedmessyexample}

\end{figure}


\subsection{Overview and Design Choices}
Utilizing LLMs for software model completion gives rise to several challenges addressed by \rag{}: how to provide context, such as domain knowledge, to the LLM, how to serialize software models, and how to deal with limited context\footnote{Software models can become huge compared to the limited number of tokens that can be given to a LLM.}?

Regarding context, we opt for \ragLong{}, and compare the approach to fine-tuning in one of our experiments. The next important design decision is that we do not work on the software models directly but on the \scg{s}, described in \secref{sec:background}.
The basic idea is that \scg{} completions can be straight forwardly interpreted as model completions (i.e., generating a new ``added'' node corresponds to adding a new model element to the model). Working with the concept of a \scg{} has several advantages: 
First, we do not have to work with the entire software model representation, but we can focus on slices of the models around recently changed elements.
This is one tactic of dealing with the common problem of the limited context of a LLM. For example, in our running example, the entire (serialized) UML \mmodel{} is huge and would not fit in the context of contemporary LLMs.

Second, \scg{} completions also include attribute changes and deletions of model elements and are not limited to the creation of new model elements.
\changedFinalRevision{
\rag{} is capable of suggesting semantically appropriate changes, such as renaming an attribute or altering the type of an attribute. Additionally, it recommends specific attribute values that are beyond predefined options, for example, values for string type attributes.
Although alternative representations besides \scg{} can influence the outcome, choosing \scg{} was a deliberate design decision we made. }{}

\begin{figure}[htbp]
	 		\centering
			\includegraphics[width=1.0\columnwidth]{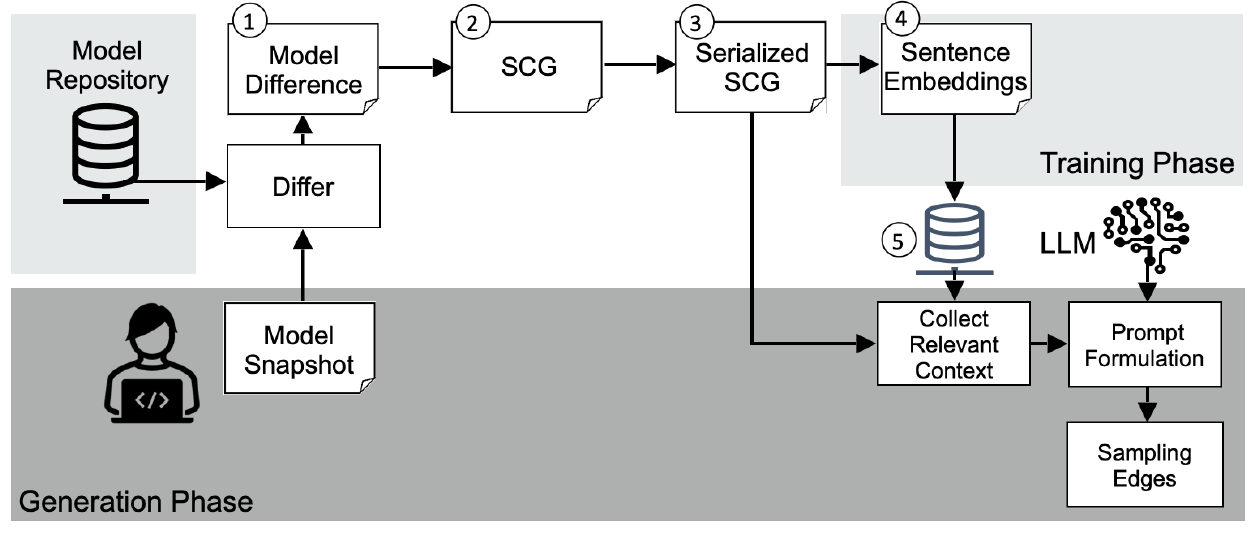}

			\caption{Overview of \rag{}.}
			\label{fig:pipelineapproach}
\end{figure}

An overview of the approach is depicted in \figref{fig:pipelineapproach},  the computation of model differences (\figref{fig:pipelineapproach}, \circled{1}) and \scg{}s (\figref{fig:pipelineapproach}, \circled{2}) is explained in \secref{sec:background}. Their serialization will be addressed in the next subsection.
\changedFinalRevision{
Based on the terminology in \secref{sec:background}, the formalization of our approach~\rag{} is given in the \suppweb{}.}{}

\subsection{Pre-processing}

Both \trainingPhase{} and \generationPhase{} work on serializations of \scg{s}. We describe how these serializations are derived based on the example given in \figref{fig:motivatingexample}.
Input to this procedure are two (successive) revisions of a model; output is a serialization of their \scg{s}. 
These revisions can originate either from the model the user is working on (in the generation phase) or from our training data.

In the first step, a model difference is computed for each pair of successive revisions of a model (\figref{fig:pipelineapproach}, \circled{1}). 
Regarding our running example in \circleda{} of \figref{fig:motivatingexample}, we also highlighted these model differences by color, that is, ``added'' model elements are depicted in green. 
From this model difference, we compute a (partial) \scg{} (see \defref{definition_scg} and \figref{fig:pipelineapproach}, \circled{2}).
Finally, the \scg{} is serialized as a list of edges (\figref{fig:pipelineapproach}, \circled{3}).
To this end, we defined a graph serialization, called \emph{\edgel{}}, for directed labeled graphs. 
\figref{fig:serializedmessyexample} presents the prompt generated from our approach alongside the corresponding response, which was retrieved via API access to ChatGPT. It also shows an example of this graph serialization (e.g., last part of the prompt), which contains all kinds of attribute information. It can quickly become verbose and noisy in real-world examples.
Common formats such as the GraphML\footnote{\url{http://graphml.graphdrawing.org}} are less suitable for LLMs, since they list vertices before edges. 
This requires guessing all nodes first -- added, deleted, and preserved -- before generating edges. 
\changedFinalRevision{}{This process is impractical, as it forces the model to specify nodes without considering their connections. In contrast, we aim to generate nodes and edges simultaneously, which aligns better with how LLMs process and interpret natural language--extending it sequentially, rather than inserting elements retroactively.}

\subsection{\TrainingPhase{}}
The \emph{input} to the \trainingPhase{} is a set of serialized \scg{} components
. The \emph{output} is a (vector) store of serializations with a key for retrieval (\figref{fig:pipelineapproach}, \circled{5}). We retrieve relevant \scg{s} from model repositories by utilizing a \emph{similarity search} based on sentence embeddings~\cite{reimers2019sentence}. The serializations are stored in a vector database together with their sentence embedding (\figref{fig:pipelineapproach}, \circled{4} and \circled{5}).

\subsection{\GenerationPhase{}}

The \emph{input} to the \generationPhase{} is a set of serialized \scg{} components capturing the difference of a new model snapshot (i.e., local changes) and the previous model revision ($m_1 \stackrel{\varepsilon}{\to} m_{2}$), as well as the vector store from the \trainingPhase{}.
The \emph{output} is a (list of) completion(s) in the form of \edgel{} serializations, which are suggested to the user after being parsed (an example is given in \figref{fig:serializedmessyexample}, at the bottom under 'Response'). 


\textbf{Retrieval.} 
The vector store is queried for \scg{} serializations via a similarity-based retrieval.
Note that, in our case the retrieved context can be interpreted as \emph{few-shot examples}, because we retrieve complete \scg{s}, that is, completed partial \scg{s} from the history. The few-shot samples from \figref{fig:serializedmessyexample} are detailed in the \suppweb{}. 
To ensure a diversity of samples, we use a procedure similar to maximum marginal relevance~\cite{Carbonell1998MMR}, explained in detail in the \suppweb{}. As few-shot samples, we select up to \maxfewshot{} serialized \scg{s}; we investigate the dependency on the number of few-shot samples in \secref{sec:evaluation}.

\textbf{Prompt formulation.} 
The prompt (input to the LLM) used by our approach consists of an instruction at the beginning, followed by the few-shot samples retrieved from the vector store (joined via a separation token), and finally the (partial)-\scg{} serialization is concatenated (see \figref{fig:serializedmessyexample}).


\textbf{Sampling new edges.} 
We can sample multiple completion candidates from the LLM by using a beam search or by instructing the LLM to generate multiple edges. Details of the edge sampling are given in the \suppweb{}.



\subsection{Implementation}
We have implemented the computation of model differences and \scg{s} on top of the \textsc{Eclipse Modeling Framework}~\cite{steinberg2008}, using \textsc{SiDiff}~\cite{schmidt2008constructing} for matching and diffing. The other parts are implemented in \textsc{Python3}, mainly utilizing \textsc{NetworkX}\footnote{\url{https://networkx.org}} for handling graphs. We use \textsc{LangChain}\footnote{\url{https://python.langchain.com}\hfill} for the handling of language models and \ragLong{}.
We use the \textsc{all-MiniLM-L6-v2}\footnote{\url{https://huggingface.co/sentence-transformers/all-MiniLM-L6-v2}} language model for the sentence embeddings since it performed well in preliminary experiments. As vector store, we use \textsc{ChromaDb}\footnote{\url{https://www.trychroma.com}}.
As language model, we use \textsc{GPT-4} (version 0613), since it performed best in preliminary experiments. We use a dedicated deployment of OpenAI on Microsoft Azure that is certified for the classification level of the industrial data. 



\section{Evaluation}\label{sec:evaluation}

We evaluate to what extent our approach is able to derive structurally and semantically correct \ceo{s} from the software model history. This includes, in particular, their applicability in industrial scenarios.
%
\changedFinalRevision{
We aim at a systematic evaluation of LLMs for model completion in a controlled setting. 
This allows us to concentrate on the core effectiveness of LLM technology, while controlling for confounding factors such as tool use and human aspects (e.g., UX design facets). }{} This is also the reason why, at this stage, conducting a user study settled in a specific application context would be not opportune (but needs to follow at a later stage).
However, by applying our approach to a real-world context at our industry partner, who expressed clear interest in and demand for this technology, we establish a solid methodological and empirical foundation, before considering the development of sophisticated and potentially costly tools. 

\subsection{Research Questions}\label{sec:rqs}
To understand the merits of language models for model completion, we want to answer the following research questions:

\begin{tcolorbox}[arc=0pt,outer arc=0pt, boxrule=0pt, top=2pt, bottom=2pt, left=2pt, right=2pt, breakable,  sharpish corners,enhanced, drop lifted shadow]
\textbf{\generalRQ:} \textit{To what extent can pre-trained language models and \ragLong{} be used for the completion of software \mbox{models}?}
\end{tcolorbox}
Clearly, a general pre-trained language model is typically not aware of the syntax and domain-specific semantics of the \scg{} serializations \emph{per se}. This includes the definition of the graph serialization format, the definition of \scg{s}, the \mmodel{}, and the domain-specific semantics of the software models not already encoded in the \mmodel{}. 
For example, a generated completion might be invalid according to the \mmodel{}, (e.g., invalid combination of edge, source, and target node labels) or could even result in an invalid directed labeled graph serialization (e.g., they do not adhere to the \edgel{} format).

\begin{tcolorbox}[arc=0pt,outer arc=0pt, boxrule=0pt, top=2pt, bottom=2pt, left=2pt, right=2pt, breakable,  sharpish corners,enhanced, drop lifted shadow]
\textbf{\semanticretrievalRQ:} \textit{
What influence does semantic retrieval have on the performance of \rag{}?}
\end{tcolorbox}
As motivated in \secref{sec:approach}, providing context that is semantically close to a to-be-completed change could improve the correctness of \ragLong{}. We therefore want to understand the influence of the similarity-based retrieval on model completion. That is, we want to compare semantic retrieval and random retrieval of few-shot examples and to analyze the influence of the number of few-shot examples.


\changedFinalRevision{
\begin{tcolorbox}[arc=0pt,outer arc=0pt, boxrule=0pt, top=2pt, bottom=2pt, left=2pt, right=2pt, breakable,  sharpish corners,enhanced, drop lifted shadow]
\textbf{\baselineRQ:} \textit{How does \rag{} compare against the state of the art (Chaaben et al.~\cite{chaaben2023towards})?}
\end{tcolorbox} 
We evaluate the accuracy of our proposed approach, \rag{}, by comparing it to the closely related work of Chaaben et al.~\cite{chaaben2023towards}, which we use as a baseline. Their study focuses on few-shot learning to suggest new model elements, providing the same unrelated, few-shot examples independently of the current model to be completed.
Our investigation centers on the prediction improvements that can be realized by providing semantically similar examples from the model history as context to the LLM for the model completion task. 

}{}

\begin{tcolorbox}[arc=0pt,outer arc=0pt, boxrule=0pt, top=2pt, bottom=2pt, left=3pt, right=3pt, breakable,  sharpish corners,enhanced, drop lifted shadow]
\textbf{\limitationsRQ:} \textit{What are limitations of using LLMs for model completion in a real-world setting?}
\end{tcolorbox}
While quantitative results provide insights into the merits of LLMs  on model completion, we also want to investigate when and why model completion fails. From simple examples and simulated changes it is hardly possible to make assertions for real-world changes. We therefore take a closer look at a sample set from \emph{real-world changes}. From our observations, we will derive research gaps and hypotheses for future research.



\begin{tcolorbox}[arc=0pt,outer arc=0pt, boxrule=0pt, top=2pt, bottom=2pt, left=2pt, right=2pt, breakable,  sharpish corners,enhanced, drop lifted shadow]
\textbf{\finetuningRQ:} \textit{What insights can be gained when comparing domain-specific fine-tuning to our retrieval-based approach \rag{}?}
\end{tcolorbox}
An alternative to \ragLong{} is domain-specific fine-tuning. We explore its viability, considering dataset properties and training specifics (e.g., epochs and base LLM).

\subsection{Datasets}\label{sec:dataset}
To answer our research questions, we make use of three datasets, balancing internal and external validity.
Basic statistics about the datasets are given in \tblref{tbl:datasets}. 

\begin{table}[htbp]
\caption{Basic statistics for the datasets. Model size is measured in terms of the number model elements. Changes include added, deleted, and modified model elements.}
\begin{adjustbox}{center}
\footnotesize
\begin{tabular}{rrrrrr}
\toprule
Dataset        & No.\  & No.\  & Avg.\  & Avg.\ No.\         & Public\\
               & Models            & Revisions               & Model Size         & Changes &       \\

\midrule
\textsc{Industry}          &  8   & 159 & 11\,365 & 50\,340  & No  \\
\textsc{RepairVision}          &  42  & 3\,139 & 685 & 70  & Yes   \\
\textsc{Synthetic}          &  24  & 360 & 5\,402 & 564  &  Yes \\
\bottomrule
\end{tabular}
\end{adjustbox}

\label{tbl:datasets}
\end{table}

\textbf{\textsc{Industry} Dataset.}
We have extracted this dataset from a repository of \textsc{SysML} models in \textsc{MagicDraw}\footnote{\textsc{MagicDraw} is a modeling tool commonly used in industries for UML and SysML (system modeling).} for a train control software used by a large product line of trains of our industry partner. The dataset stems from an industry collaboration, where we tackle several challenges related to the management of large industrial software product lines. The model for the train control software comprises several submodels, such as drive and brake control, interior lightning, exterior lightning, sanitary facilities, HVAC, etc. \changedFinalRevision{In a preprocessing step, we have removed confidential information (e.g., the models contain requirement owner information and other personal information of involved engineers). }{} 
The models themselves as well as the average number of changes between revisions in this dataset are large (cf. Table \ref{tbl:datasets}). The large number of changes originates from many attributes changes, such as renamings, and typically long time periods between two revisions.

The \textsc{Industry} dataset 
with its domain-specific and project-specific concepts
helps to understand to what extent we can use LLMs for software model completion in a complex, real-world setting. It allows us to assess the effectiveness in navigating the noisy, complex, and often irregular nature of real-world data -- a critical aspect often overlooked in existing research.

\textbf{\textsc{RepairVision} Dataset.}
The \textsc{RepairVision} \cite{ohrndorf2021history,ohrndorf2021b} dataset is a public dataset\footnote{\url{https://repairvision.github.io/evaluation}} of real-world open-source models, containing histories of 21 \textsc{Ecore} repositories, such as UML2 or BPMN2.
The \textsc{RepairVision} dataset plays a crucial role in our evaluation in assessing how effectively LLMs can be employed for software model completion in real-world settings. Similar to the \textsc{Industry} dataset, the serialized change graphs in this dataset can become verbose and noisy and reflect the difficulties of real-world model completion (see \figref{fig:serializedmessyexample}). Its public availability facilitates reproducibility, comparability, and public accessibility, fundamental aspects that ensure our research can be examined and extended by others.

\textbf{\textsc{Synthetic} Ecore Dataset.}
With the first two datasets, we aimed at external validity and a real-world setting. At the same time, we had only little control over potentially influential factors of the dataset impairing internal validity. 
To obtain a dataset for which we can control several properties of the model repositories, 
we simulated the evolution of a software model similar to Tinnes et al.~\cite{Tinnes2021}: We used a \mmodel{} that resembles a simple component model \changedFinalRevision{(as used in modelling system architecture) with {\small \sffamily components}, {\small \sffamily implementations}, {\small \sffamily ports}, {\small \sffamily connectors}, and {\small \sffamily requirements}. } {} 
Some predefined \eo{s} have been randomly applied to a revision of a software model to obtain a new revision of the software model.  
This way we were able to control the number of edit operations that are applied per model revision (i.e., 11, 31, 51, 81) and the number of model revisions in one dataset (i.e., 10 or 20). We furthermore randomly applied perturbations. That is, with a certain probability (i.e., 0\%, 50\%, 100\%), we slightly modified the edit operation by a successive application of an additional \eo{} that overlaps with the original \eo{}. 
The repositories in this dataset contain only changes at the type level, that is, we do no include attributes or changes thereof. 
The \textsc{Synthetic} dataset gives us more control over several properties of a model repository, allowing us to specifically understand how fine-tuning is affected by the properties of the model repositories, this way increasing internal validity.



\subsection{Operationalization} \label{sec:operationalization}

We conduct \nbExperiments{} experiments, one per research question. 
For all significance tests, we use a significance level of $\alpha=0.05$.

\textbf{\generalExp (\generalRQ):} To answer \generalRQ, we preprocess all three datasets from \secref{sec:dataset} and generate a collection with training (\trainratio{}) and testing samples (\testratio{}), \changedFinalRevision{more specifically \scg{}s, to ensure a systematic evaluation.
We then select\footnote{The selection procedure is explained in detail in the \suppweb{}.} between \numbersamples{} samples, depending on the dataset from the testing set and}{}, for each, we select between 1 to \maxfewshot{} few-shot samples from the training set.
The reason to choose between \numbersamples{} samples is (1) to obtain a sample set of a manageable size that we can manually analyze and that induces acceptable costs for the LLM usage and (2) to obtain a large enough set to draw conclusions.

First, we analyze the correctness of the generated completions \changedFinalRevision{with respect to the ground truth. A \scg{} contains a change that actually occurred in the modeling history. From the change graph, we randomly remove edges to obtain a partial change graph, with the full change graph being the corresponding ground truth.
This approach improves over previous methods that involves arbitrarily removing elements from a static snapshot. By focusing on model histories, we create a realistic setting, selecting subsets of changes that have actually occurred in real-world scenarios. }{}
\changedFinalRevision{We consider different levels of correctness: \emph{Structural correctness} ensures that the graph structure is correct, with properly directed, sourced, and targeted nodes. \emph{Change structure correctness} builds on this by additionally requiring correct types of changes to the model, such as whether elements should be modified, added, or removed. Lastly, \emph{type structure correctness} demands further an exactly correct 'type' and 'changetype'. An illustrative example for these types is given in \figref{fig:serializedmessyexample} under 'response'. We automatically check the format, structural correctness, change semantics, and type correctness for all datasets. }{}

For the \textsc{industry} dataset, we additionally manually evaluate the generated completions to also check for \emph{semantic} correctness. 
\changedFinalRevision{
In our manual analysis of \emph{semantic correctness}, a solution was deemed correct if the LLM's proposed completion matched the ground truth in meaning and purpose. This check cannot be automated due to the extensive use of natural language in our data 
and application-specific identifiers (e.g., user-chosen attribute names). For example, in \figref{fig:serializedmessyexample}, 
naming a new operation 'getExtension' or 'getExt' is a matter of preference, while their semantic meaning is the same.
}{}
\changedFinalRevision{We addressed potential errors and bias in our manual analysis by having two of the authors independently evaluate the proposed solutions. Any mismatches in their evaluations were discussed, and a consensus was reached on the correct interpretation. 
}{}
For the base LLM, we use \textsc{GPT-4}\footnote{
Note that we experimented with several LLMs from the \textsc{GPT} family of models and also observed changes in the specific model's performance over 
time~\cite{chen2023chatgpts}.
At the time of execution, GPT-4 using a small introductory prompt that explains the tasks (see \suppweb)
was performing best on a small test set, and we therefore fixed the LLM in \rag{} to \textsc{GPT-4}. 
} (version 0613) in a dedicated Azure deployment to complete our prompts.


\textbf{\semanticretrievalExp (\semanticretrievalRQ):}
In \semanticretrievalRQ, we investigate whether the correctness (from correct format to semantic correctness) depends on the number of few-shot samples.
For the \textsc{Industry} dataset, we have the information on whether a few-shot sample's change is of a similar class as the test \scg{}. 
We also investigate how this affects correctness, that is, whether the similarity-based retrieval in \rag{} affects the correctness of completions.
To this end, we compare semantic sampling with few-shot samples that have been randomly retrieved from the training data.
We evaluate this for semantic correctness. For this reason, and also to reduce the LLMs usage costs, we perform this analysis only for the \textsc{Industry} dataset. 

\changedFinalRevision{\textbf{\baselineExp (\baselineRQ):} To address \baselineRQ, we selected the publicly available \textsc{Revision} dataset. This selection not only enhances reproducibility but also allows for comparisons with future methodologies, such that ongoing research advancements can be directly compared to our \rag{} and the work by Chaaben et al.~\cite{chaaben2023towards}. }{}
\changedFinalRevision{
Their approach recommends new classes, their associations, and attributes. Accordingly, the present experiment specifically targets these aspects. We excluded samples that did not fall into these categories, resulting in \numbersamplesforbaselinecomparison{} test examples from the \textsc{Revision} dataset for comparison.
To replicate the approach introduced by Chaaben et al., which we denote as a \textsc{Baseline}, we use their few-shot examples, serialization of concepts, and incorporate the partial models similarly into the prompt. Further details are available in the \suppweb{}. We query \textsc{GPT-3} (text-davinci-002) several times and suggest the most frequently occurring concept. 
}{}


\textbf{\limitationsExp (\limitationsRQ):} We answer \limitationsRQ\ by manually investigating completions that have been generated in the first experiment for the \textsc{Industry} dataset. 
We go through all prompt and completion pairs and identify common patterns where the model completion works well or does not, and we aim at interfering causes that led to the results. Since this 
analysis is time-consuming, we focus on the \textsc{Industry} dataset -- a domain- and project-specific, real-world dataset.
We report on the identified strengths and weaknesses of the approach -- given this real-world scenario -- and point to research gaps and formulate hypotheses for future research and improvements. 

\textbf{\finetuningExp (\finetuningRQ):} 
To investigate whether fine-tuning is a viable alternative to few-shot prompting (see \generalExpwithout), we fine-tune models from the \textsc{GPT} family of language models on the \textsc{Synthetic} dataset. The reasons why we restrict this analysis to the \textsc{Synthetic} datasets are manifold: The main reason is that we want to understand \emph{how} the performance of the fine-tuning approach depends on various properties of the dataset in a controlled setting. Furthermore, we have a limited budget for this experiment, and fine-tuning is costly. 
We also control for the number of fine-tuning epochs and the base language model used for the fine-tuning.
For every repository of the dataset, we split the data into training set (90\%) and testing set (10\%), and we use the test set to report on the performance of the completion task. The fine-tuning of the models optimizes the average token accuracy\footnote{
At the time of experiment execution, evaluating with any self-defined test metrics was not possible using the fine-tuning APIs provided by OpenAI.
This metric is not aware of any specifics of the dataset, and even a single wrong token in a serialization can produce a
syntactically wrong serialization, while the token accuracy for the incorrect completion would still be high.}.
%
To compare the \ragLong{} to fine-tuning, we run both for the same test samples.
For the few-shot training samples, we also use the same training samples used to fine-tune the language models. 
We assess the correctness with regard to the ground truth. Due to the unique characteristics of the \textsc{Synthetic} dataset,
the ground truth correctness is defined by the graph structure, change structure, and type structure.

\subsection{Results}

\textbf{\generalExp (\generalRQ):}
Addressing \generalRQ, which explores the extent to which pre-trained LLMs and \ragLong{} can be utilized for software model completion, our findings on the correctness of \rag{} are detailed in Table \ref{tbl:correctness}.

We list the different \emph{levels of correctness} for all datasets.
We see that more than 90\% of the completions have a correct format and even more than 76\% of 
completions are type correct, that is, completed edges have the right source and target nodes, and type and the types of the source and target node are correct. 
Even at a semantic level, 62\% of the generated completions are correct for the \textsc{Industry} dataset. For the \textsc{Synthetic} dataset type correctness is equivalent to semantic correctness. Consequently 86\% of the results are correct for this dataset.
\begin{table}[htbp]

\caption{Different levels of correctness in percent (\%) of the entire test set for all three datasets.}
   
    \begin{adjustbox}{center}
    \footnotesize
    \begin{tabularx}{\columnwidth}{ccccccc}
    \toprule
    &                 &  & Change    & Type        &   & Total\\
    
    Dataset & Format & Structure &  Structure &  Structure & Semantic  &  Count    \\

    \midrule
    \textsc{Industry}           &  92.62   & 86.89 & 78.69 & 76.23  & 62.30  & 122\\
    \textsc{RepairVision}       &  91.86   & 84.62 & 84.16 & 76.92  & --      & 221\\
    \textsc{Synthetic}          &  99.05   & 86.19 & 86.19 & 86.19  & --      & 210\\
    \bottomrule
    \end{tabularx}
    \end{adjustbox}
  
    \label{tbl:correctness}
    \end{table}


\textbf{\semanticretrievalExp (\semanticretrievalRQ):}
Regarding the relationship between the number of few-shot samples and correctness,
we conducted a (one-sided) Mann-Whitney-U test for the overall and type/semantic correct distributions over the number few-shot samples.
For every dataset, we do not find any significant relationship between the number of few-shot samples and correctness (smallest $p$-value is $0.2$ for the type correctness of the \textsc{RepairVision} dataset). 
Furthermore, we find that test samples where a similar class of changes is among the few-shot samples perform significantly better than overall correctness ($p=0.0289$ using a Mann-Whitney-U test, $p=0.0227$ using a binomial test).
Finally, we find that similarity-based retrieval performs significantly better than random retrieval for type correctness ($p< 10^{-9}$, using a binomial test) as well as for semantic correctness ($p<0.0038$ by a binomial test\footnote{For semantic correctness, we rely on the fact that the number of semantically correct samples is smaller than the number of type correct samples. Thus, we are able to compute an upper bound for the $p$-value using the type correct random retrieval samples.}).

\begin{table}[htbp]
\caption{Different levels of correctness in percent (\%) of \rag{} and random retrieval on the \textsc{Industry} dataset.
}

\begin{adjustbox}{center}
\begin{tabular}{ccccccc}
\toprule
&                 &  & Change    & Type        &   & Total \\
Approach & Format & Structure &  Structure &  Structure & Semantic  &  (Count)    \\ 
\midrule
\rag{}          &  92.62   & 86.89 & 78.69 & 76.23  & 62.30  & 122\\
\textsc{Random}  &  84.43   &  79.51& 52.46 & 50.00  & --      & 122\\
\bottomrule
\end{tabular}
\end{adjustbox}
\label{tbl:correctnesscomparedtobaseline}
\end{table}


\changedFinalRevision{
\textbf{\baselineExp (\baselineRQ):} To obtain a clear picture of the pros and cons of \rag{}, \textsc{Baseline} and random retrieval, we independently report the accuracy of the correct concepts (classes) and the correct association. We further split correct concepts in correct type (``Same Class'' in Table \ref{tbl:allbaselines}) and correct name (see \suppweb{}, for details). }{}
\changedFinalRevision{
We perform binomial tests (our random baseline against \rag{} and Chaaben et al.) to compare the effectiveness of our approach. We found that, in all cases, \rag{} performs significantly better than \textsc{Random}, which, in turn, performs even significantly better than  \textsc{Baseline} (Table~\ref{tbl:allbaselines}). 
}{}

\begin{table}[htbp]
\vspace{-0.25em}
\caption{
Different levels of correctness (\%) of \rag{}, random retrieval, and \textsc{Baseline} on the \textsc{Revision} dataset.
}

\begin{adjustbox}{center}
  \begin{threeparttable}[b]

\begin{tabular}{ccccc}
\toprule
Approach & Same Class & Same Name & Same Concept &  Same Assoc. \\
\midrule
\rag{} & 94.1\tnote{**} & 96.1\tnote{**} & 94.1\tnote{**} & 80.4\tnote{*} \\
\textsc{Random}  & 78.4 & 80.4 & 76.5 & 68.6 \\
\textsc{Baseline}& 21.6\tnote{**} & \phantom{1}9.8\tnote{**} & \phantom{1}9.8\tnote{**} & \phantom{1}7.8\tnote{**}\\
\bottomrule
\end{tabular}
  \begin{flushright}
      \scriptsize
      (**: $p < 0.01$, *: $p<0.05$)
    \end{flushright}
 \end{threeparttable}
\end{adjustbox}

\label{tbl:allbaselines}
\end{table}


\textbf{\limitationsExp (\limitationsRQ):}
To better understand when and why the \ragLong{} succeeds or fails when completing software models, we separate our analysis here in two parts---successful completions and unsuccessful ones.


\textit{Reoccurring patterns (success):} Several of the successful completions follow repeating completion patterns. 
For example, there is a move refactoring, where a {\small \sffamily package} declaration with type definitions is moved from one {\small \sffamily package} to another {\small \sffamily package}. Since this happened quite often in the past repository histories, the correct new parent {\small \sffamily package} could be deduced, even though this {\small \sffamily package} is not yet part of the incomplete test sample. 

\textit{Complex refactorings (success):} Furthermore, more complex refactorings have also been be completed correctly, for example, a redesign of a whole-part decomposition including {\small \sffamily packages} and \textsc{SysML} {\small \sffamily block} definitions has been correctly performed. Similarly, we find correctly completed refactorings dealing with inheritance (of port types).

\textit{Project-specific concepts (success):} Even project-specific concepts, such as a special kind of tagging concept to mark software components as ``frozen'', are correctly inferred from the few-shot examples or co-changes of {\small \sffamily components} are correctly identified, likewise. 

\textit{No memorization (success):} We also observe correct handling of structure in non-trivial cases. For example, correct combinations of source and target node ids are generated can not be observed in the few-shot examples. 

\textit{Noise (success):} We also observe that the language model is able to infer concepts among noise, that is, unrelated changes. For example, there are correctly completed instances of the ``add {\small \sffamily interface block} and {\small \sffamily type} reference'' concept where similar few-shot samples are only present with lots of entangled changes.

Regarding unsuccessful cases, we observe two main reasons for failure: incorrect structure and incorrect semantic.

\textit{Structural conflicts (failure):} For incorrect structure, we find examples where conflicts occur because a node with the same node id is already present. Furthermore, sometimes (correct) model elements or packages are added to the incorrect parent package (in most cases, we see a tendency of the LLM to ``flatten'' hierarchies).

\textit{Structure incorrect (failure):} There are several instances where correct edge, source, and target node types are generated but their ids, and consequently the structure, is incorrect. 

\textit{Semantics wrong b/c copy\&paste (failure):} One cause for incorrect semantic completions is that parts of few-shot samples are incorrectly copied and pasted. \changedFinalRevision{This typically occurs when the LLM lacks sufficient context to generate the correct completion, leading it to mistakenly copy and paste segments from the provided examples.}{This usually happens when there is no context available to make the right completion.}

\textit{Semantics wrong b/c unknown evolution/missing context (failure):} For example, in the case of functional project-specific evolution, it might be hard to ``guess'' the right completion without further knowledge, or 
the semantic retrieval might fail to retrieve instances of the correct change pattern. Interestingly, in some of these cases, the LLM is ``guessing well but not perfect'' (e.g., added {\small \sffamily subsystem} instead of {\small \sffamily external subsystem}).



\textit{Conceivable but unobserved evolution (failure):}
Another interesting instance of incorrect semantic completion is a completion where a comment (in German) should be removed but instead a comment (in English) has been added. In the project, there were many renamings from German to English and, in this case, a future change has been correctly anticipated.

 \textbf{\finetuningExp (\finetuningRQ):}
To compare our \ragLong{-based} to fine-tuning, we perform an analysis at the token level, and we also compare the completions on a graph-structural and semantic level.
At the token level, we find an average token accuracy of \averageTokenAccuracy{}, with a minimum of \minTokenAccuracy{}, and a maximum of \maxTokenAccuracy{} on our test data sets (10\% test ratio).
We can observe strong correlation of the average token accuracy with the number of fine-tuning epochs. Also, larger models perform better with respect to the average token accuracy. Regarding the repository properties, we only find significant negative correlations with the perturbation probability. That is, more diverse repositories are typically harder for the model completion using fine-tuning. Exact numbers are given in our \suppweb. 
When comparing the distributions of the edges removed in the \scg{} for incorrect and correct completions, we see that the average number of removed edges for the incorrect (i.e., no exact match) completions ($5.78$) is significantly larger than the average number of removed edges for the correct ones ($2.94$). Similarly, we find a significant relationship for the distributions of the total \scg{} size ($14.89$ for the incorrect completions, and $6.39$ for correct completions).
Accuracies of the comparison of our approach to the fine-tuning approach are given in Table \ref{tbl:comparingisoex4}.
%
%
%
%
 %
\begin{table}[htbp]

\caption{Different levels of correctness in percent (\%) for fine-tuned models compared to the retrieval-based approach in multi-edge software model completion on \textsc{Synthetic}.}

\vspace{-0.5
em}
\begin{adjustbox}{center}
\footnotesize
\begin{tabularx}{\columnwidth}{XXcc}
  \toprule
Dataset & Method & Correct edge(s)  & Exact match 
 \\ 
  \midrule
\datasetadacompletion   & \rag{}  &88.52 & 39.34 \\ 
 & \ada{} & 88.33 & 56.67 \\ 
\datasetcuriecompletion & \rag{} & 86.00 & 37.00 \\ 
  &\curie{} & 90.05  & 64.68 \\ 
   \bottomrule
\end{tabularx}
\end{adjustbox}

\label{tbl:comparingisoex4}

\end{table}

We conducted a Mann-Whitney-U test to compare the performance of \ragLong{} and the fine-tuned \curie{} and \ada{} models from the GPT-3 family. In terms of producing, at least, one correct edge, neither fine-tuning nor \ragLong{} exhibit statistical significance in outperforming the other.
In terms of exact matches,
\ada{} ($p=0.0290$) and \curie{} ($ p< 10^{-7}$) outperform \ragLong{}. 
Regarding exact matches, the impact of different sampling methods used in fine-tuning and \rag{} becomes substantial
(algorithms are provided in the \suppweb{}).
While \rag{} often produces more edges than required, the sampling procedure used with the fine-tuning models is more conservative.

\subsection{Discussion}


Overall, we find that both \rag{} and fine-tuning of LLMs
are promising approaches for model completion, and the general inference capabilities of LLMs are useful, can handle noisy contexts, and provide real-time capabilities. 
We will next discuss the results, outline hypotheses for potential future research, and describe threats to validity in Section~\ref{sec:threats}.

\textbf{\generalRQ:} In the \generalExpwithout, we observed promising correctness values across all datasets.
Not only are more than 90\% of completions correct w.r.t.\ the serialization format, but we also find \emph{more than 62\% of semantically correct completions in the real-world industrial setting}. This indicates that \ragLong{} is a promising technique for model completion. 
Token processing times fall within the millisecond range, and time required for semantic retrieval is negligible, even for larger models.
The approach’s real-time capability is significant given the stepwise model completion use case.

\textbf{\semanticretrievalRQ:}
In the \ragLong{} setting, we do not find any significant relationship between the number of few-shot samples and correctness. 
We find that similarity-based retrieval boosts the correctness of the approach and that
it significantly performs better if a similar relevant change---following a similar pattern---is available in the context. 
It also worthwhile mentioning that real-world datasets are typically biased with respect to the change pattern, and semantic retrieval can avoid sampling from large but irrelevant change pattern.

\changedFinalRevision{\textbf{\baselineRQ:} We have observed that, in all instances where new elements with associations are recommended, \rag{} consistently outperforms random retrieval and Chaaben et al.~\cite{chaaben2023towards}. These results reinforce our findings from \generalRQ, namely that leveraging LLMs with \ragLong{} represents a viable approach for model completion.
}{}

\textbf{\limitationsRQ:}\label{sec:discussionRQ2}
We have seen that our approach can be used to provide completions that are correct to a large extent for simple reoccurring patterns but also more complex refactorings. Even project-specific concepts can be deduced from few-shot examples. In many cases, generated edges are also structurally correct. The general inference capabilities of LLMs are useful, for example, in dealing with concepts for which there are few or no similar examples. Furthermore, also with noise \ragLong{} often provides correct completions.
\changedFinalRevision{
Regarding usefulness of the completions, 
our manual analysis reveals that many of the completions appear useful for the modeler. For example, \rag{} was able to perform a translation of several German comments to English, because the engineering language of the project has been changed. Furthermore, \rag{} was able to complete project-specific refactorings.
}{}
For a further investigation of these observations, we formulate the following hypothesis.
\begin{tcolorbox}[arc=0pt,outer arc=0pt, boxrule=0pt, top=1pt, bottom=1pt, left=2pt, right=2pt, breakable,  sharpish corners,enhanced, drop lifted shadow]

\textbf{Hypothesis 1:} 
LLMs and \ragLong{} are able to handle noisy training examples, leverage (domain) knowledge from pre-training\changedFinalRevision{, adapt to project-specific concepts, and provide useful software model completions}{}. 
\end{tcolorbox}

We found completions that are incorrect from a structural viewpoint as well as incorrect from a semantic viewpoint. 
As for structurally incorrect completions, we identified cases where existing node ids are incorrectly reused, where incorrect (containment) hierarchies would have been created, or where completed edges are correct from a type perspective but do not connect the right nodes. It is worth further investigating how these structural deficiencies could be overcome, in particular, given that LLMs are designed for sequential input, not for graph inputs. 
This leaves us with the following hypothesis.

\begin{tcolorbox}[arc=0pt,outer arc=0pt, boxrule=0pt, top=1pt, bottom=1pt, left=2pt, right=2pt, breakable,  sharpish corners,enhanced, drop lifted shadow]

\textbf{Hypothesis 2:} 
Conceivable remedies for the structural deficiencies include fine-tuning of LLMs, 
combining graph neural networks -- designed for graph-like input -- with LLMs, providing multiple different graph serialization orders, 
or a positional encoding that reflects the graph-like nature of the \scg{} serializations. 
\end{tcolorbox}

\vskip 1ex
Regarding semantics, we found incorrect completions that were related to a lack of (domain) knowledge in 
the pre-trained model or the few-shot examples, respectively. For example, we found cases of functional evolution 
where the language model is missing (domain) knowledge or requirements, 
or cases of a refactoring without any relevant few-shot sample. 
\changedFinalRevision{ 
We further identified cases where a conceivable completion has been generated but was not the one from the ground truth.}{}

\begin{tcolorbox}[arc=0pt,outer arc=0pt, boxrule=0pt, top=1pt, bottom=1pt, left=2pt, right=2pt, breakable,  sharpish corners,enhanced, drop lifted shadow]

\textbf{Hypothesis 3:} 
Conceivable remedies for the semantic deficiencies include strategies to further fuse the approach with context knowledge 
(e.g., fine-tuning, providing requirements, or task context in the prompt, leveraging other project data in repositories etc.). 
Furthermore, providing a list of recommendations may cure some identified deficiencies.
\end{tcolorbox}


\textbf{\finetuningRQ:}
We found that a more fine-tuning epochs are beneficial for the average token accuracy. More diverse repositories increase the difficulty for the software model completion.
The larger the \scg{} and the more edges we omit for the completion, the higher the probability of an incorrect completion.
The reason that fine-tuning has a higher exact match accuracy is more due to the edge sampling algorithm than to the method itself: When analyzing the percentage of correct edges, it becomes clear that we cannot conclude that one approach outperforms the other. 
Instead, we hypothesize a strong dependency on the edge sampling procedure, which deserves further investigation. While the \ragLong{} often generated more edges than necessary, the sampling procedure used with the fine-tuned models from the GPT-3 family takes a more conservative approach, prioritizing the generation of edges with high confidence.






\textbf{Comparison to code completion.}
Note that LLMs for source code completions
show similar results to our findings in Experiment~1 and~5, ranging from~29\% for perfect prediction of entire code blocks to~69\% for a few tokens in a single code statement~\cite{ciniselli22}. 
Drawing a direct comparison between code and model completion is not straightforward, though.

\subsection{Threats to Validity}\label{sec:threats}
With respect to construct validity, we made several design choices that may not be able to leverage the entire 
potential of LLMs for software model completion, including our definition of \scg{s}, the serialization of the \scg{},
the strategy of how to provide domain knowledge to the language model, and the choice of the base LLM.


To increase internal validity, we incorporated the \textsc{Synthetic} Ecore Dataset into our experiments, controlling for 
properties of software model repositories.
Still, we were not always able to completely isolate every factor in our
experiments. For example, fine-tuning and few-shot learning use different edge samplings. This is due to
the API that we used to access the language models. In future research, an ablation study 
for the design choices in the algorithms shall be performed. 
\changedFinalRevision{To address the potential variability that LLMs may exhibit, we checked and confirmed that the completions were stable. }{}

Regarding external validity, we included two real-world datasets (\textsc{RepairVision} and \textsc{Industry}), and we study real-world change scenarios from the observed history in these repositories. 
We have chosen our test samples to be small enough to perform manual semantic analysis, but large enough to draw conclusions.
\changedFinalRevision{To minimize costly manual checks, our semantic analysis was confined to our most challenging dataset, the \textsc{Industry} dataset. Extending the analysis to the other datasets would enhance validity. However, we are confident that, having analyzed hundreds of samples, we have struck a reasonable compromise. }{}
We are therefore certain that our results have an acceptable degree of generalizability for the current state of research. In any case, user studies 
shall investigate the usefulness of our completions in practice.
Investigating merits of LLMs for model completion is an emerging topic, and many questions are open. Still, our results set a lower bound for the potential
of LLMs in this area, with promising results, insights, and hypotheses for further research.

\section{Conclusion}

We presented and investigated an approach to software model completion based on \ragLong{}, \rag{}, and compared it to fine-tuning during our evaluation.
Our experiments on a simulated, a public, open source \textsc{Ecore}, and an industrial \textsc{SysML} dataset for a train control software product line
show that, indeed, LLMs are a promising technology for software model completion. The real-time capability of our approach is especially beneficial for stepwise model completion, highlighting its practical utility. 
We achieved a semantic correctness in a real-world industry setting of \semanticCorrectIndustryPercent{}, which is comparable to earlier results with LLMs for source code completion.
Further investigation revealed that similarity-based retrieval significantly enhances the correctness of model completions and that fine-tuning is a viable alternative to \ragLong{}.
All in all, the general inference capabilities of LLMs are beneficial, particularly in dealing with concepts for which only  scarce or even no analogous examples are provided. 
We have identified concrete causes for the technology to fail and formulated corresponding hypotheses for future research.
Of utmost importance for future research is to compare technology, such as graph neural networks, that has been designed for processing graph-like data (e.g., our \scg{s}),
especially for structural aspects of software model completion. Also, marrying approaches that are strong for structural aspects, and LLMs,
that are typically strong for semantic aspects of model completion is worth further investigation.

\section{Data Availability}
\label{sec:data_availability}
We provided all data (excluding the \textsc{Industry} dataset) and the Python code of our approach on a \suppwebrepo{}. 
\changedFinalRevision{We cannot include the \textsc{Industry} dataset, because it contains sensitive data, including intellectual property of products on the market. }{}
We provide R scripts and Jupyter Notebooks to replicate our statistical evaluation. 

\def\bibfont{\footnotesize}
\bibliographystyle{plain}
\bibliography{bib}

\begin{thebibliography}{10}

\bibitem{adhikari2023simima}
Bhisma Adhikari, Eric~J Rapos, and Matthew Stephan.
\newblock Simima: a virtual simulink intelligent modeling assistant: Simulink intelligent modeling assistance through machine learning and model clones.
\newblock {\em Software and Systems Modeling}, pages 1--28, 2023.

\bibitem{agt2018domore}
Henning Agt-Rickauer, Ralf-Detlef Kutsche, and Harald Sack.
\newblock Domore--a recommender system for domain modeling.
\newblock In {\em Proceedings of the International Conference on Model-Driven Engineering and Software Development}, volume~1, pages 71--82. Set{\'u}bal: SciTePress, 2018.

\bibitem{agt2019automated}
Henning Agt-Rickauer, Ralf-Detlef Kutsche, and Harald Sack.
\newblock Automated recommendation of related model elements for domain models.
\newblock In {\em Model-Driven Engineering and Software Development: 6th International Conference, MODELSWARD 2018, Funchal, Madeira, Portugal, January 22-24, 2018, Revised Selected Papers 6}, pages 134--158. Springer, 2019.

\bibitem{ahmad2023towards}
Aakash Ahmad, Muhammad Waseem, Peng Liang, Mahdi Fahmideh, Mst~Shamima Aktar, and Tommi Mikkonen.
\newblock Towards human-bot collaborative software architecting with chatgpt.
\newblock In {\em Proceedings of the 27th International Conference on Evaluation and Assessment in Software Engineering}, pages 279--285, 2023.

\bibitem{ahmad2021unified}
Wasi~Uddin Ahmad, Saikat Chakraborty, Baishakhi Ray, and Kai-Wei Chang.
\newblock Unified pre-training for program understanding and generation.
\newblock {\em arXiv preprint}, 2021.

\bibitem{ahmed2022few}
Toufique Ahmed and Premkumar Devanbu.
\newblock Few-shot training llms for project-specific code-summarization.
\newblock In {\em Proceedings of the International Conference on Automated Software Engineering (ASE)}, pages 1--5, 2022.

\bibitem{almonte22}
Lissette Almonte, Esther Guerra, Iv{\'{a}}n Cantador, and Juan de~Lara.
\newblock Recommender systems in model-driven engineering.
\newblock {\em Software and System Modelling}, 21(1):249--280, 2022.

\bibitem{arendt2010henshin}
Thorsten Arendt, Enrico Biermann, Stefan Jurack, Christian Krause, and Gabriele Taentzer.
\newblock {Henshin: Advanced concepts and tools for in-place {EMF} model transformations}.
\newblock In {\em Proceedings of the International Conference on Model Driven Engineering Languages and Systems (MODELS)}, pages 121--135. Springer, 2010.

\bibitem{Biermann2012}
Enrico Biermann, Claudia Ermel, and Gabriele Taentzer.
\newblock {Formal foundation of consistent EMF model transformations by algebraic graph transformation}.
\newblock {\em Software and Systems Modeling}, 11(2):227--250, 2012.

\bibitem{brun2008model}
C{\'e}dric Brun and Alfonso Pierantonio.
\newblock Model differences in the eclipse modeling framework.
\newblock {\em UPGRADE, The European Journal for the Informatics Professional}, 9(2):29--34, 2008.

\bibitem{bucchiarone2020grandchallenges}
Antonio Bucchiarone, Jordi Cabot, Richard~F Paige, and Alfonso Pierantonio.
\newblock Grand challenges in model-driven engineering: an analysis of the state of the research.
\newblock {\em Software and Systems Modeling}, 19:5--13, 2020.

\bibitem{burgueno2021nlp}
Loli Burgue{\~n}o, Robert Claris{\'o}, S{\'e}bastien G{\'e}rard, Shuai Li, and Jordi Cabot.
\newblock {An NLP-based architecture for the autocompletion of partial domain models}.
\newblock In {\em Proceedings of the International Conference on Advanced Information Systems Engineering}, pages 91--106. Springer, 2021.

\bibitem{cabot2018cognifying}
Jordi Cabot, Robert Claris{\'o}, Marco Brambilla, and S{\'e}bastien G{\'e}rard.
\newblock Cognifying model-driven software engineering.
\newblock In {\em Software Technologies: Applications and Foundations}, pages 154--160. Springer, 2018.

\bibitem{camara2023assessment}
Javier C{\'a}mara, Javier Troya, Lola Burgue{\~n}o, and Antonio Vallecillo.
\newblock On the assessment of generative ai in modeling tasks: an experience report with chatgpt and uml.
\newblock {\em Software and Systems Modeling}, pages 1--13, 2023.

\bibitem{Carbonell1998MMR}
Jaime Carbonell and Jade Goldstein.
\newblock The use of mmr, diversity-based reranking for reordering documents and producing summaries.
\newblock In {\em Proceedings of the International Conference on Research and Development in Information Retrieval}, page 335–336, New York, NY, USA, 1998. ACM.

\bibitem{chaaben2023towards}
Meriem~Ben Chaaben, Lola Burgue{\~n}o, and Houari Sahraoui.
\newblock Towards using few-shot prompt learning for automating model completion.
\newblock In {\em Proceedings of the International Conference on Software Engineering: New Ideas and Emerging Results (ICSE-NIER)}, pages 7--12. IEEE, 2023.

\bibitem{chen2023chatgpts}
Lingjiao Chen, Matei Zaharia, and James Zou.
\newblock How is chatgpt's behavior changing over time?
\newblock {\em arXiv}, 2023.

\bibitem{chen2021codex}
Mark Chen, Jerry Tworek, Heewoo Jun, Qiming Yuan, Henrique Ponde de~Oliveira Pinto, Jared Kaplan, Harri Edwards, Yuri Burda, Nicholas Joseph, Greg Brockman, et~al.
\newblock Evaluating large language models trained on code.
\newblock {\em arXiv preprint}, 2021.

\bibitem{tsigkanos23}
Tsigkanos Christos, Rani Pooja, Müller Sebastian, and Kehrer Timo.
\newblock Large language models: the next frontier for variable discovery within metamorphic testing?
\newblock In {\em Proceedings of the International Conference on Software Analysis, Evolution and Reengineering}. {IEEE}, 2023.

\bibitem{ciniselli22}
Matteo Ciniselli, Nathan Cooper, Luca Pascarella, Antonio Mastropaolo, Emad Aghajani, Denys Poshyvanyk, Massimiliano Di~Penta, and Gabriele Bavota.
\newblock An empirical study on the usage of transformer models for code completion.
\newblock {\em Transactions on Software Engineering}, 48(12):4818--4837, 2022.

\bibitem{dabney2004mastering}
James~B Dabney and Thomas~L Harman.
\newblock {\em Mastering simulink}, volume 230.
\newblock Pearson/Prentice Hall Upper Saddle River, 2004.

\bibitem{damasceno2021quality}
Carlos Diego~Nascimento Damasceno and Daniel Str{\"u}ber.
\newblock Quality guidelines for research artifacts in model-driven engineering.
\newblock In {\em Proceedings of the International Conference on Model Driven Engineering Languages and Systems (MODELS)}, pages 285--296. IEEE, 2021.

\bibitem{deng2016recommendation}
Shuiguang Deng, Dongjing Wang, Ying Li, Bin Cao, Jianwei Yin, Zhaohui Wu, and Mengchu Zhou.
\newblock A recommendation system to facilitate business process modeling.
\newblock {\em IEEE transactions on cybernetics}, 47(6):1380--1394, 2016.

\bibitem{di2023memorec}
Juri Di~Rocco, Davide Di~Ruscio, Claudio Di~Sipio, Phuong~T Nguyen, and Alfonso Pierantonio.
\newblock Memorec: a recommender system for assisting modelers in specifying metamodels.
\newblock {\em Software and Systems Modeling}, 22(1):203--223, 2023.

\bibitem{di2022finding}
Juri Di~Rocco, Claudio Di~Sipio, Phuong~T Nguyen, Davide Di~Ruscio, and Alfonso Pierantonio.
\newblock Finding with nemo: a recommender system to forecast the next modeling operations.
\newblock In {\em Proceedings of the 25th International Conference on Model Driven Engineering Languages and Systems}, pages 154--164, 2022.

\bibitem{di2023morgan}
Claudio Di~Sipio, Juri Di~Rocco, Davide Di~Ruscio, and Phuong~T Nguyen.
\newblock Morgan: a modeling recommender system based on graph kernel.
\newblock {\em Software and Systems Modeling}, pages 1--23, 2023.

\bibitem{Ehrig2004}
Hartmut Ehrig, Ulrike Prange, and Gabriele Taentzer.
\newblock {Fundamental theory for typed attributed graph transformation}.
\newblock In {\em International Conference on Graph Transformation (ICGT)}, pages 161--177. Springer, 2004.

\bibitem{elkamel2016uml}
Akil Elkamel, Mariem Gzara, and Han{\^e}ne Ben-Abdallah.
\newblock An uml class recommender system for software design.
\newblock In {\em Proceedings of the International Conference of Computer Systems and Applications (AICCSA)}, pages 1--8. IEEE, 2016.

\bibitem{feng2020codebert}
Zhangyin Feng, Daya Guo, Duyu Tang, Nan Duan, Xiaocheng Feng, Ming Gong, Linjun Shou, Bing Qin, Ting Liu, Daxin Jiang, et~al.
\newblock Codebert: A pre-trained model for programming and natural languages.
\newblock {\em arXiv preprint}, 2020.

\bibitem{gamma1995design}
Erich Gamma, Ralph Johnson, Richard Helm, Ralph~E Johnson, and John Vlissides.
\newblock {\em Design patterns: elements of reusable object-oriented software}.
\newblock Prentice Hall, 1995.

\bibitem{gomes2023dome}
Anderson Gomes and Paulo Henrique~M Maia.
\newblock Dome: An architecture for domain model evolution at runtime using nlp.
\newblock In {\em Proceedings of the XXXVII Brazilian Symposium on Software Engineering}, pages 186--195, 2023.

\bibitem{heinemann2012facilitating}
Lars Heinemann.
\newblock Facilitating reuse in model-based development with context-dependent model element recommendations.
\newblock In {\em 2012 Third International Workshop on Recommendation Systems for Software Engineering (RSSE)}, pages 16--20. IEEE, 2012.

\bibitem{huang2021short}
Ningyuan~Teresa Huang and Soledad Villar.
\newblock A short tutorial on the weisfeiler-lehman test and its variants.
\newblock In {\em International Conference on Acoustics, Speech and Signal Processing (ICASSP)}, pages 8533--8537. IEEE, 2021.

\bibitem{iec61131plc}
IEC.
\newblock Programmable controllers - part 3: Programming languages.
\newblock Technical report, DIN/EN/IEC 61131, 2014.

\bibitem{iovino2020model}
Ludovico Iovino, Angela Barriga~Rodriguez, Adrian Rutle, and Rogardt Heldal.
\newblock Model repair with quality-based reinforcement learning.
\newblock 2020.

\bibitem{jesse2023large}
Kevin Jesse, Toufique Ahmed, Premkumar~T Devanbu, and Emily Morgan.
\newblock Large language models and simple, stupid bugs.
\newblock {\em arXiv}, 2023.

\bibitem{Kehrer2015}
Timo Kehrer.
\newblock {\em {Calculation and Propagation of Model Changes based on User-Level Edit Operations: A Foundation for Version and Variant Management in Model-Driven Engineering}}.
\newblock PhD thesis, University of Siegen, 2015.

\bibitem{Kehrer2017}
Timo Kehrer, Abdullah~M Alshanqiti, and Reiko Heckel.
\newblock {Automatic inference of rule-based specifications of complex in-place model transformations}.
\newblock In {\em Proceedings of the International Conference on Model Transformations (ICMT)}, pages 92--107. Springer, 2017.

\bibitem{Kehrer2016}
Timo Kehrer, Gabriele Taentzer, Michaela Rindt, and Udo Kelter.
\newblock {Automatically deriving the specification of model editing operations from meta-models}.
\newblock In {\em Proceedings of the International Conference on Model Transformations (ICMT)}, volume 9765, pages 173--188, 2016.

\bibitem{kogel2017recommender}
Stefan K{\"o}gel.
\newblock Recommender system for model driven software development.
\newblock In {\em Proceedings of the 2017 11th Joint Meeting on Foundations of Software Engineering}, pages 1026--1029, 2017.

\bibitem{kogel2016automatic}
Stefan K{\"o}gel, Raffaela Groner, and Matthias Tichy.
\newblock Automatic change recommendation of models and meta models based on change histories.
\newblock In {\em ME@ MoDELS}, pages 14--19, 2016.

\bibitem{kruchten1995}
Philippe~B Kruchten.
\newblock The 4+ 1 view model of architecture.
\newblock {\em IEEE software}, 12(6):42--50, 1995.

\bibitem{kuschke2017rapmod}
Tobias Kuschke and Patrick M{\"a}der.
\newblock Rapmod—in situ auto-completion for graphical models.
\newblock In {\em Proceedings of the International Conference on Software Engineering (ICSE): Companion Proceedings}, pages 303--304. IEEE, 2017.

\bibitem{kuschke2013recommending}
Tobias Kuschke, Patrick M{\"a}der, and Patrick Rempel.
\newblock Recommending auto-completions for software modeling activities.
\newblock In {\em International conference on model driven engineering languages and systems}, pages 170--186. Springer, 2013.

\bibitem{langer2013posteriori}
Philip Langer, Manuel Wimmer, Petra Brosch, Markus Herrmannsd{\"o}rfer, Martina Seidl, Konrad Wieland, and Gerti Kappel.
\newblock A posteriori operation detection in evolving software models.
\newblock {\em Journal of Systems and Software}, 86(2):551--566, 2013.

\bibitem{li2013efficient}
Ying Li, Bin Cao, Lida Xu, Jianwei Yin, Shuiguang Deng, Yuyu Yin, and Zhaohui Wu.
\newblock An efficient recommendation method for improving business process modeling.
\newblock {\em IEEE Transactions on Industrial Informatics}, 10(1):502--513, 2013.

\bibitem{lopez2022modelset}
Jos{\'e} Antonio~Hern{\'a}ndez L{\'o}pez, Javier~Luis C{\'a}novas~Izquierdo, and Jes{\'u}s~S{\'a}nchez Cuadrado.
\newblock Modelset: a dataset for machine learning in model-driven engineering.
\newblock {\em Software and Systems Modeling}, pages 1--20, 2022.

\bibitem{minas2009}
Steffen Mazanek and Mark Minas.
\newblock Business process models as a showcase for syntax-based assistance in diagram editors.
\newblock In Andy Sch{\"u}rr and Bran Selic, editors, {\em Model Driven Engineering Languages and Systems}, pages 322--336, Berlin, Heidelberg, 2009. Springer Berlin Heidelberg.

\bibitem{mazanek2009generating}
Steffen Mazanek and Mark Minas.
\newblock {Generating correctness-preserving editing operations for diagram editors}.
\newblock {\em Electronic Communication of the European Association of Software Science and Technology}, 18, 2009.

\bibitem{maeder2021}
Patrick Mäder, Tobias Kuschke, and Mario Janke.
\newblock Reactive auto-completion of modeling activities.
\newblock {\em Transactions on Software Engineering}, 47(7):1431--1451, 2021.

\bibitem{nassar2017rule}
Nebras Nassar, Hendrik Radke, and Thorsten Arendt.
\newblock Rule-based repair of emf models: An automated interactive approach.
\newblock In {\em Theory and Practice of Model Transformation: 10th International Conference, ICMT 2017, Held as Part of STAF 2017, Marburg, Germany, July 17-18, 2017, Proceedings 10}, pages 171--181. Springer, 2017.

\bibitem{neubauer2017automated}
Patrick Neubauer, Robert Bill, Tanja Mayerhofer, and Manuel Wimmer.
\newblock Automated generation of consistency-achieving model editors.
\newblock In {\em 2017 IEEE 24th International Conference on Software Analysis, Evolution and Reengineering (SANER)}, pages 127--137. IEEE, 2017.

\bibitem{ohrndorf2021b}
Manuel Ohrndorf, Christopher Pietsch, Udo Kelter, Lars Grunske, and Timo Kehrer.
\newblock History-based model repair recommendations.
\newblock {\em Transactions of Software Engineering Methodology (TOSEM)}, 30(2), 2021.

\bibitem{ohrndorf2021history}
Manuel Ohrndorf, Christopher Pietsch, Udo Kelter, and Timo Kehrer.
\newblock {ReVision: A tool for history-based model repair recommendations}.
\newblock In {\em Proceedings of the International Conference on Software Engineering (ICSE): Companion Proceedings}, pages 105--108. ACM, 2018.

\bibitem{omg2013mof}
OMG.
\newblock {OMG Meta Object Facility (MOF) Core Specification, Version 2.4.1}.
\newblock Technical report, Object Management Group, June 2013.

\bibitem{UML2017}
OMG.
\newblock Unified modeling language ({UML}) version 2.5.1.
\newblock Standard, Object Management Group, December 2017.

\bibitem{SysML2019}
OMG.
\newblock Omg sysml v. 1.6.
\newblock Standard, Object Management Group, December 2019.

\bibitem{Polanyi1958}
Michael Polanyi.
\newblock {\em {Personal Knowledge: Towards a Post Critical Philosophy}}.
\newblock University of Chicago Press, 1958.

\bibitem{reimers2019sentence}
Nils Reimers and Iryna Gurevych.
\newblock Sentence-bert: Sentence embeddings using siamese bert-networks.
\newblock {\em arXiv preprint}, 2019.

\bibitem{robles2023reflection}
Gregorio Robles, Michel~RV Chaudron, Rodi Jolak, and Regina Hebig.
\newblock A reflection on the impact of model mining from github.
\newblock {\em Information and Software Technology}, 164:107317, 2023.

\bibitem{RodriguesDaSilva2015}
Alberto {Rodrigues Da Silva}.
\newblock {Model-driven engineering: A survey supported by the unified conceptual model}.
\newblock {\em Computer Languages, Systems and Structures}, 43:139--155, 2015.

\bibitem{samoaa2022systematic}
Hazem~Peter Samoaa, Firas Bayram, Pasquale Salza, and Philipp Leitner.
\newblock A systematic mapping study of source code representation for deep learning in software engineering.
\newblock {\em IET Software}, 2022.

\bibitem{schmidt2008constructing}
Maik Schmidt and Tilman Gloetzner.
\newblock {Constructing difference tools for models using the SiDiff framework}.
\newblock In {\em Proceedings of the International Conference on Software Engineering (ICSE): Companion Proceedings}, pages 947--948. {ACM/IEEE}, 2008.

\bibitem{sen10completion}
Sagar Sen, Benoit Baudry, and Hans Vangheluwe.
\newblock Towards domain-specific model editors with automatic model completion.
\newblock {\em Simulation}, 86(2):109--126, 2010.

\bibitem{Sobania2021GPvsGPT}
Dominik Sobania, Martin Briesch, and Franz Rothlauf.
\newblock Choose your programming copilot: {A} comparison of the program synthesis performance of github copilot and genetic programming.
\newblock {\em CoRR}, abs/2111.07875, 2021.

\bibitem{steimann2013}
Friedrich Steimann and Bastian Ulke.
\newblock Generic model assist.
\newblock In Ana Moreira, Bernhard Sch{\"a}tz, Jeff Gray, Antonio Vallecillo, and Peter Clarke, editors, {\em Proceedings of the International Conference on Model Driven Engineering Languages and Systems (MODELS)}, pages 18--34. Springer Berlin Heidelberg, 2013.

\bibitem{steinberg2008}
Dave Steinberg, Frank Budinsky, Ed~Merks, and Marcelo Paternostro.
\newblock {\em EMF: eclipse modeling framework}.
\newblock Pearson Education, 2008.

\bibitem{stephan2019towards}
Matthew Stephan.
\newblock Towards a cognizant virtual software modeling assistant using model clones.
\newblock In {\em 2019 IEEE/ACM 41st International Conference on Software Engineering: New Ideas and Emerging Results (ICSE-NIER)}, pages 21--24. IEEE, 2019.

\bibitem{stephan2013survey}
Matthew Stephan and James~R Cordy.
\newblock A survey of model comparison approaches and applications.
\newblock In {\em Proceedings of the International Conference on Model-Driven Engineering and Software Development (MODELSWARD)}, pages 265--277, 2013.

\bibitem{tinnes2023mining}
Christof Tinnes, Timo Kehrer, Mitchell Joblin, Uwe Hohenstein, Andreas Biesdorf, and Sven Apel.
\newblock Mining domain-specific edit operations from model repositories with applications to semantic lifting of model differences and change profiling.
\newblock {\em Automated Software Engineering}, 30(2):17, 2023.

\bibitem{Tinnes2021}
Christof Tinnes, Timo Kehrer, Joblin. Mitchell, Uwe Hohenstein, Andreas Biesdorf, and Sven Apel.
\newblock Learning domain-specific edit operations from model repositories with frequent subgraph mining.
\newblock In {\em Proceedings of the International Conference on Automated Software Engineering (ASE)}. ACM/IEEE, 2021.

\bibitem{visser2007model}
Arie Van~Deursen, Eelco Visser, and Jos Warmer.
\newblock {Model-driven software evolution: A research agenda}.
\newblock {\em Technical Report Series TUD-SERG-2007-006.}, 2007.

\bibitem{Varro2006}
D{\'{a}}niel Varr{\'{o}}.
\newblock {Model transformation by example}.
\newblock In {\em Proceedings of the International Conference on Model Driven Engineering Languages and Systems (MODELS)}, pages 410--424. Springer, 2006.

\bibitem{vaswani2017attention}
Ashish Vaswani, Noam Shazeer, Niki Parmar, Jakob Uszkoreit, Llion Jones, Aidan~N Gomez, {\L}ukasz Kaiser, and Illia Polosukhin.
\newblock Attention is all you need.
\newblock {\em Advances in Neural Information Processing Systems}, 30, 2017.

\bibitem{wang2022no}
Chaozheng Wang, Yuanhang Yang, Cuiyun Gao, Yun Peng, Hongyu Zhang, and Michael~R Lyu.
\newblock No more fine-tuning? an experimental evaluation of prompt tuning in code intelligence.
\newblock In {\em Proceedings of the European Software Engineering Conference and Symposium on the Foundations of Software Engineering}, pages 382--394, 2022.

\bibitem{weyssow2022recommending}
Martin Weyssow, Houari Sahraoui, and Eugene Syriani.
\newblock Recommending metamodel concepts during modeling activities with pre-trained language models.
\newblock {\em Software and Systems Modeling}, 21(3):1071--1089, 2022.

\bibitem{xu2022systematic}
Frank~F Xu, Uri Alon, Graham Neubig, and Vincent~Josua Hellendoorn.
\newblock A systematic evaluation of large language models of code.
\newblock In {\em Proceedings of the International Symposium on Machine Programming}, pages 1--10, 2022.

\bibitem{Xu2022LLMCode}
Frank~F. Xu, Uri Alon, Graham Neubig, and Vincent~Josua Hellendoorn.
\newblock A systematic evaluation of large language models of code.
\newblock In {\em Proceedings of the International Symposium on Machine Programming}, page 1–10. ACM, 2022.

\bibitem{zhao2021natural}
Liping Zhao, Waad Alhoshan, Alessio Ferrari, Keletso~J Letsholo, Muideen~A Ajagbe, Erol-Valeriu Chioasca, and Riza~T Batista-Navarro.
\newblock Natural language processing for requirements engineering: a systematic mapping study.
\newblock {\em ACM Computing Surveys (CSUR)}, 54(3):1--41, 2021.

\end{thebibliography}

\newpage
\begin{appendix}
\label{app:supplementary}

\subsection{Sampling of Experiment Samples}
\changedFinalRevision{
To get a manageable (e.g., we want to perform a manual analysis for correctness and need to keep LLM usage cost acceptable) -- but still large enough -- sample for our experiments, our aim is to sample around 100--200 examples for each of the datasets (i.e., \textsc{Synthetic}, \textsc{Repairvision}, and \textsc{Industry}). Every dataset consists of several projects (e.g., submodels in the case of the \textsc{Industry} dataset), and we ensure that they are represented with the same distribution in our sample. We draw at least 200 examples. Since we ensure that every project is included, at least, once, even if it is very small, this leaves us with \numbersamplessynthetic{} samples for the \textsc{Synthetic} dataset, \numbersamplesrevision{} samples for the \textsc{RepairVision} dataset, and \numbersamplesindustry{} samples for the \textsc{Industry} dataset. For every project in the dataset we used a diversity sampling strategy (cf. \secref{sec:semanticretrieval}) to obtain a diverse range of samples. 
For the \textsc{Industry} dataset, since we perform a manual analysis of semantic correctness there, we further want to reduce the size of the sample, without harming diversity too much.
We could have just randomly (or with the procedure above) down-sampled, for example, to 100 samples. Instead, since we generally observe a strong homogeneity in the real-world datasets (i.e., many repeating patterns in \textsc{Industry} and \textsc{RepairVision} datasets), we wanted to ensure that we do not decrease the heterogeneity.
We therefore decided to further select samples from industry with a more controlled procedure:

We further examine the prompt and completion pairs to classify the changes into semantic clusters that we defined. We skimmed the dataset once and came up with a list of change patterns, for example, ``interface added between components''. We agreed on a final list of patterns (the first and second author of the paper agreed on categories for the examples from the initially drawn \numbersamplesindustry{} examples).
We then recorded whether the training samples contain a change that falls into the same class. We then only included samples that are unique according to the number of training examples in the prompt, the class of the change that we assigned, and whether there is a similar change in the training samples or not (which has also been decided with the help of the patterns we defined for the changes). This leaves \numberindustrysamples{} samples for the model completion task on the \textsc{industry} dataset.}{}

\subsection{Formalization}\label{sec:approach-formalization}

\changedFinalRevision{
In addition to the terminology from our formalization in \secref{sec:background}, we will define further operators and then formalize our approach~\rag{} on top:
We define a serialization operator $s\colon \mathcal{G} \to \Sigma ^{*}$, where $\Sigma ^{*}$ is the set of all strings. This operator $s$ takes a (partial) \scg{} $g \in \mathcal{G}$ and returns a serialization for this graph (the detailed serialization is given in \secref{sec:serialization}). Furthermore, we define $\operatorname{r}\colon \Sigma ^{*} \to {\Sigma ^{*}}^k$, which, given a (partial) serialized \scg{} $s(g) \in \Sigma ^{*}$, retrieves $k$  ``similar'' serialized \scg{s} from a (vector) database. We define the prompt operator $\operatorname{prompt}\colon \Sigma ^{*} \times {\Sigma^{*}}^k \to \Sigma ^{*}$, which, given a tuple $(s(g), \operatorname{r}(s(g))$ of the (partial) \scg{} and the retrieved similar serialized \scg{s}, constructs the final prompt (described in detail in \secref{sec:candidateGeneration}). Given the prompt instruction $i \coloneq \operatorname{prompt((s(g), \operatorname{r}(s(g)))}$, we can generate serialized completed \scg{} candidates by sampling tokens with a LLM. That is, we sample tokens $\omega_{j+1}$ from $\mathbb{P}(\omega_{j+1}|i \, \omega_1\dots\omega_j)$, until the entropy becomes too large or a complete edge serialization has been sampled. A large entropy of the language model token probabilities can be seen as an indicator\footnote{assuming a well-calibrated LLM} for a high uncertainty of further tokens. We denote this candidate generation by
\begin{alignat*}{1}
\operatorname{cg_{LLM}}\colon \Sigma^{*} &\to \Sigma^{*}\\
\operatorname{cg_{LLM}}(i) &\mapsto s(g) \, \omega_{1} \dots \omega_{J},
\end{alignat*}
where $J$ is the total number of sampled tokens. By $s(g) \, \omega_{1}$, we denote the string concatenation of $s(g)$ and $\omega_{1}$, and likewise for the rest of this expression.
Finally, we parse $s(g) \, \omega_{1} \dots \omega_{J}$ as a graph (if possible), and interpret it as a completed \scg{}, which represents a model transformation $\gamma$.
It can happen (although it rarly happens in practise as can be seen in the evaluation section of this paper) that the completed string does not represent our simple change graph serialization format. In this case, we record this failure and consider the model completion as failed.
Identifying the parsed graph with the corresponding edit operation, we denote this parsing operator by $s^{-1}\colon \Sigma ^{*} \to \mathcal{E}$. With this notation, the entire model completion approach, \rag{}, can be formalized by

\begin{alignat*}{2}
        \operatorname{C_{\rag{}}} \colon \mathcal{T} &\to \mathcal{T} &&\\
        (m_1, m_2) &\mapsto \pi(m_1,\,&&s^{-1}\,\circ\,\operatorname{cg_{LLM}}\,\circ\,\operatorname{prompt}\,\\
        \,&\,&&\circ\,\text{id} \times \operatorname{r}\,\circ\,s\,\circ\,\operatorname{SCG} (m_1, m_2)),
\end{alignat*}
where $\pi$ is the application operator and $\operatorname{SCG}$ the simple change graph operator defined in \secref{sec:background}.

}{}

\subsection{Implementation Details}
\subsubsection{Realization of the operator $\operatorname{SCG}$: Simple Change Graphs and Their Labels}\label{sec:app-labels}
In Definition \ref{definition_scg}, we defined \scg{s} as subgraphs of a difference graph, which is a labeled graph.
In this section, we explain in more detail, how we derive the labels from the models and the change graph\changedFinalRevision{, that is, the realization of the operator $\operatorname{SCG}$.}{} 

We assume a simplified \mmodel{}, in which we have {\small \sffamily classes} that carry a {\small \sffamily name}, that is, the {\small \sffamily type} of a model element. A {\small \sffamily class} has {\small \sffamily attributes} that have a {\small \sffamily attribute name} and {\small \sffamily attribute value}, and  {\small \sffamily references} that have a {\small \sffamily reference type}. 

For a given model, we then use this simplified \mmodel{} to derive a labeled graph (cf. Definition \ref{def:labeled_graph}): we map {\small \sffamily objects} (i.e., instances of a {\small \sffamily class}) to a node of the labeled graph and instances of {\small \sffamily references} to edges. By this, we ensure that our graph representation is structurally equivalent to an abstract syntax graph of the model (difference).
Nodes and edges in the graph carry a label. For nodes, this label is a JSON representation of the object. It has a attribute {\small \sffamily type} with its value equal to the  {\small \sffamily name} of the {\small \sffamily class} the  {\small \sffamily object} is an instance of. It also contains all {\small \sffamily attributes} with their values for the given {\small \sffamily object}(assuming we can serialize the attribute). More concretely, the {\small \sffamily attributes} are a contained as a  nested JSON inside the node label JSON with JSON attribute names equal to the {\small \sffamily attribute name} and JSON value given by the {\small \sffamily attribute value}. Finally, for the edge labels, we use a JSON that has a JSON attribute {\small \sffamily type} with value equal to the {\small \sffamily reference type}.

Next, for the difference graph, we simply add to each node and each edge another JSON attribute  {\small \sffamily changeType}, with value equal to  {\small \sffamily Add},  {\small \sffamily Preserve}, or  {\small \sffamily Remove}, depending on the change type in the difference graph. For modified attributes, we add another node attached to the necessarily preserved  {\small \sffamily object} with a JSON label indicating the attribute value \emph{before} and \emph{after} the change.
Since a \scg{} is a subgraph of the difference graph, this construction also defines the labels of the \scg{}.

Note that in some cases (e.g., to check for type correctness), we can simply remove attribute information from our labels, thus obtaining a graph that has only information about the type structure. We use this graph, for example, to check for type correctness of model completions. 
Furthermore, this graph without attribute information can also be helpful for other use cases, where we are only interested in the type structure, for example, in change pattern mining use cases, or if we want to define a reusable template for \eo{s}. 

Now that we know how to construct a labeled graph for a given model difference, we will next see how we serialize these labeled graphs.

\subsubsection{Realization of the operator $s$: Serialization Format}\label{sec:serialization}
In this section, we explain our serialization format for graphs, called \edgel{}, which will be part of the prompt being send to the LLM (cf. \figref{fig:serializedmessyexample}).
\changedFinalRevision{In the language of \secref{sec:approach-formalization}, this section it about the implementation of the operator $s$. }{}

The serialization of a graph starts with a header line (indicating an id of the graph).  
\begin{lstlisting}[frame=single, basicstyle=\sffamily\scriptsize, breakindent=0pt]
t # <graph_id>
\end{lstlisting}
After the header, all edges of the graph are serialized edge-by-edge, where one edge will correspond to one line in the serialization format.
An edge is represented by one line of the following format:

\begin{lstlisting}[frame=single, basicstyle=\sffamily\scriptsize, breakindent=0pt]
e <src_id> <tgt_id> <src_label> <tgt_label> <edge_label>
\end{lstlisting}

Here, \lstinline[basicstyle=\sffamily\small]{<src_label>}, \lstinline[basicstyle=\sffamily\small]{<tgt_label>}, and \lstinline[basicstyle=\sffamily\small]{<edge_label>} are the labels of the labeled graph corresponding to the \scg{} (cf. \secref{sec:app-labels}), and \lstinline[basicstyle=\sffamily\small]{<src_id>} and \lstinline[basicstyle=\sffamily\small]{<tgt_id>} are identifiers for the source and target vertices of the edge, respectively.

An extract of an example \scg{} serialization is given in Listing \ref{list:excerpt}.

When we designed this serialization format, we had already the application of LLMs for model completion in mind. More common graph serialization formats start with a list of nodes and then list edges between these nodes. Instead, we define nodes implicitly, while defining edges. Therefore, node labels of already defined nodes will be duplicated in our approach. In practise, we avoid this though, by replacing an already defined node label by an \emph{empty JSON}.

Especially in the case of fine-tuning, we do want to avoid that the LLM has first to \emph{guess} the right nodes of the graph before it continues with the edges. The \edgel{} format allows for a continuous generation of edges and avoids the break between listing nodes and listing edges.


\begin{lstlisting}[frame=single, caption={An example \SCG{} in the \edgel{} format.}, basicstyle=\sffamily\scriptsize, breakindent=1pt, captionpos=b, label={list:excerpt}, breaklines=true, postbreak=\mbox{\textcolor{red}{$\hookrightarrow$}\space}]
t # 1
e 0 1 {..."add", "type":"port"} {..."add", "type":"component"} {..."add", "type":"port"}
e 0 2 {..."add", "type":"requirement"} {..."add", "type":"component"} {..."add", "type":"requirement"}
\end{lstlisting}

In a textual representation, we have to linearize also the listing of the edges, that is, we need to decide on an ordering of the edges of the graph. In our case,
the order of edges for this serialization is determined using a depth-first search, since it proved to perform best in a pilot study. Nevertheless, other serialization strategies (or even representing the graph in more than one edge order) are conceivable and could be investigated as part of future work. 

\subsubsection{Realization of the operator $r$: Diversity Based Few-Shot Sample Retrieval}\label{sec:semanticretrieval}
\rag{} involves retrieving similar examples to the software model the user is currently working on. \changedFinalRevision{This retrieval is given by the operator $r$ in \secref{sec:approach-formalization}. }{}
To ensure diversity, typical implementations of maximum marginal relevance retrieve elements, element by element, 
and ensures maximal distance to the already existing elements. 
This can lead to below optimal samples, because samples that have already been retrieved are later not removed. In essence, typical maximum marginal relevance implementation can get stuck in local optima. 

In our sampling algorithm, the goal is the same as in maximum marginal relevance. That is, we want to select samples that are similar to a given input but the samples themselves are diverse.
We extend on maximum marginal relevance by using the following retrieval procedure: First, for a given embedding, we retrieve a given number $n$ of elements that are similar to this given embedding. We call this set $S$. Second, from $S$, we want to draw another sample of a given size $k$, that maximizes the distances between all 
elements. Initially, we draw $k$ random elements from $S$. Let's call this set $D$
Third, we choose one of these elements $e$ and replace it by an element from $(S \setminus D) \cup \{e\}$ that has maximum distance to the
$D \setminus \{e\}$. Finally, we iterate this procedure for a given number of iterations and try to choose at least one element of the initial set $D$ once.

\subsubsection{Realization of the operator $\operatorname{cg_{LLM}}\circ\operatorname{prompt}$: Candidate Generation}\label{sec:candidateGeneration}
We utilize two different tactics/algorithms to generate candidates for the software model completion. \changedFinalRevision{In the language from \secref{sec:approach-formalization}, these represent two different implementations of the operator $\operatorname{cg_{LLM}}\circ\operatorname{prompt}$}{}. In the first tactic, we keep the control over the sampling procedure and use the language model to generate the completions token-wise. We therefore use this tactic only with a ``completion-like'' interface. This tactic is more expensive, since we have to process the entire context for every token. Especially for \textsc{GPT-4}, this tactic is not feasible (without major adaptions).
For the second tactic, we utilize the LLM's capabilities to directly generate entire candidate completions. In the present study, we are using this tactic for all completions generated with \textsc{GPT-4}.

\paragraph{Beam-like Sampling Algorithm}
\label{sec:sampling}

The candidate generation works as follows (see pseudo code in Listing \ref{lst:candidateGeneration}):
The algorithm takes a set of \emph{incomplete} \eo{} candidates (in the form of serialized \scg{s}) and uses the (fine-tuned) language model to sample new edge candidates and appends them to the incomplete \eo{} candidate (Line~12). 
The sampling generates all possible extensions above a certain probability threshold.
Since we cannot guarantee that the extensions lead to a correct \edgel{} serialization, we check the syntactical correctness and reject incorrect extensions (Line~13). Furthermore, even syntactically valid extensions could be invalid according to the \mmodel{} and have to be rejected likewise (Line~14). After that, the corresponding \scg{} represents a valid \eo{} by definition. Based on a graph isomorphism test, we then filter out duplicates (Line~15). Although graph isomorphism is theoretically expensive from a computational perspective, in our setting, it is acceptable since we have only a few medium size graphs, and employ Weisfeiler-Lehman hashes~\cite{huang2021short} to speed up the comparison.  We add complete candidates to the output list (Line~19) and repeat this process until all candidates are complete (Line~9). Whether a candidate is complete is checked using several conditions such as the total probability of the candidate, a drop in the probability of a generated edge, or a generated stop token.
    
\begin{algorithm}
\caption{Pseudocode for the candidate generation.}\label{lst:candidateGeneration}
\SetCommentSty{small}

\SetKw{Function}{Function}
\SetKwFunction{GenerateCandidates}{generateCandidates}
\SetKwFunction{CheckCorrectSCG}{checkCorrectSCG}
\SetKwFunction{SampleEdges}{sampleEdges}
\SetKwFunction{CheckMetaModel}{checkMetaModel}
\SetKwFunction{Complete}{complete}
\SetKwFunction{Prune}{prune}
\SetKwFunction{Size}{size}

\SetKwData{Lm}{$\mathcal{L}$}
\SetKwData{Mm}{$\mathcal{TM}$}

\SetKwData{ExtendedEOs}{ext}
\SetKwData{IncompleteEOs}{incomplete}
\SetKwData{CompleteEOs}{complete}

\SetKwData{ExtendedEO}{$\tilde\varepsilon$}
\SetKwData{Eo}{op}
\SetKwComment{Comment}{$\triangleright~$}{}

\Function \GenerateCandidates{$\varepsilon$, \Lm, \Mm}:

\Begin{
    \KwIn{$\varepsilon$ -- given context serialization\\
    \Lm - fine-tuned language model\\
    \Mm - metamodel}
    \KwOut{$[\varepsilon_1,\dots,\varepsilon_n]$ - list of candidates}
    \IncompleteEOs $\gets$ [$\varepsilon$]\Comment*[r]{set of incomplete edit operations}
    \CompleteEOs $\gets$ [] \Comment*[r]{set of complete edit operations}
    \While{\Size{\IncompleteEOs} $>0$}{
        \ExtendedEOs $\gets$ []\Comment*[r]{set of extended edit operations}
        \ForEach{\Eo $\in$ \IncompleteEOs}{
            \ExtendedEOs $\pluseq$ \SampleEdges{\Lm, \Eo}\;
            \ExtendedEOs $\gets$ \CheckCorrectSCG{\ExtendedEOs}\;
            \ExtendedEOs $\gets$ \CheckMetaModel{\Mm, \ExtendedEOs}\;
            \ExtendedEOs $\gets$ \Prune{\ExtendedEOs, \CompleteEOs}\;
            \IncompleteEOs $\gets$ []\;
            \ForEach{\ExtendedEO $\in$ \ExtendedEOs}{
                \eIf{\Complete{\ExtendedEO}}
                {\CompleteEOs $\pluseq$ \ExtendedEO \;}
                {\IncompleteEOs $\pluseq$ \ExtendedEO \;}
            }
        }
    }
   \Return{\CompleteEOs}
}

\end{algorithm}

\paragraph{ChatModel Instruction}\label{sec:chatInstructions}
An alternative to the token-wise beam search above is to let the LLM decide when to stop.
If multiple candidates should be generated, one could sample with a certain temperature $>0$.

For our completion generation, we use the following instruction prompts:

\begin{lstlisting}[frame=single, basicstyle=\sffamily\scriptsize, breaklines=true, breakindent=0pt, caption={Single edge completion prompt.}]
You are an assistant that is given a list of change graphs in an edge format. That is, the graph is given edge by edge. The graphs are directed, labeled graphs. An edge is serialized as
"e src_id tgt_id edge_label src_label tgt_label"

Labels are dictionaries. If a node appears in more than one edge, the second time it appears it is replaced by "_" to avoid repetition. 

E.g.:
e 0 1 a b bar
e 1 2 bla _ foo

The second edge here would be equivalent to:
"e 1 2 bla bar foo"

There are some change graphs given as examples. Graphs are separated by "\n\n$$\n---\n".

The last graph in this list of graphs is not yet complete. Exactly one edge is missing. 
Your task is it, to complete the last graph by guessing the last edge. You can guess this typically by looking at the examples and trying to deduce the patterns in the examples. Give this missing edge in the format
"e src_id tgt_id edge_label src_label tgt_label". Note that the beginning "e" is already part of the prompt.
\end{lstlisting}

\begin{lstlisting}[frame=single, basicstyle=\sffamily\scriptsize, breaklines=true, breakindent=0pt, caption={Multiple edge completion prompt.}]
You are an assistant that is given a list of change graphs in an edge format. That is, the graph is given edge by edge. The graphs are directed, labeled graphs. An edge is serialized as
"e src_id tgt_id edge_label src_label tgt_label"

Labels are dictionaries or concatenations of change type and node/edge type. If a node appears in more than one edge, the second time it appears it can be replaced by "_" to avoid repetition. 

E.g.:
e 0 1 a b bar
e 1 2 bla _ foo

The second edge here would be equivalent to:
"e 1 2 bla bar foo"

There are some change graphs given as examples. Graphs are separated by "\n\n$$\n---\n".

The last graph in this list of graphs is not yet complete. Some edges are missing. 
Your task is it, to complete the last graph by guessing the missing edges. You can guess this typically by looking at the examples and trying to deduce the patterns in the examples. Give the missing edges in the format
"e src_id tgt_id edge_label src_label tgt_label". Note that the beginning "e" is already part of the prompt. After the last edge of the change graph, add two new lines.
\end{lstlisting}

\subsubsection{Realization of the operators $\pi$ and $\operatorname{SCG}$: Evaluation based on graphs in $\mathcal{G}$}
\changedFinalRevision{
Applying an \eo{} to a model $m_1$ requires some pre-conditions to be fulfilled. This includes the definition of a matching (i.e., to define the preserved model elements). As argued in \secref{sec:background}, we assume models to be valid and application conditions (pre-conditions) of \eo{s} to be fulfilled. This is the responsibility of the modeling tool. For example, in the Eclipse Modeling Framework Ecosystem, there are model transformation tools such as \textsc{Henshin}~\cite{arendt2010henshin} that can be used for the definition, verification, and application of an \eo{}. 

Given a matching and the original (left-hand side) model $m_1$ the operator $\operatorname{SCG}$ and $\pi$ are inverse to each other (i.e., $\operatorname{SCG} \, \circ \, \pi \equiv \operatorname{id}$ under the identification of $\mathcal{E}$ with the range of $\operatorname{SCG}$ in $\mathcal{G}$.

The matching defined by the matching of the incomplete \eo{} $\varepsilon$ can be reused for the matching of the completed \eo{} $\gamma \circ \varepsilon$. In case the completed \eo{} does not depend on preserved model elements that are not present in the incomplete \eo{}, the model completion will be already fully defined by a surrogate operation $C_\mathcal{G} \colon \mathcal{G} \to \mathcal{G}$. In mathematical terms, in this case the following diagram commutes.}{}

\begin{tikzcd}
\mathcal{T} \arrow[r, "\operatorname{SCG}"] \arrow[d, "C_{\rag{}}"'] & \mathcal{E} \subset \mathcal{G} \arrow[d, dashed, "C_{\mathcal{G}}"] \arrow[r, "s"] & \Sigma^{*} \arrow[d, "cg_{LLM} \circ \operatorname{prompt} \circ \operatorname{id} \times r"] \\
\mathcal{T}  & \mathcal{E} \subset \mathcal{G} \arrow[l, "\pi{(m_1,.)}"] \arrow[r, "s"'] & \Sigma^{*}
\end{tikzcd}

\changedFinalRevision{
We can therefore base our evaluation on the surrogate operator $C_{\mathcal{G}}$.
For the evaluation, we can therefore sidestep the question of defining $\pi$. 

Anyway, $\pi$ can be defined by realized by adding model elements corresponding to ``added'' nodes to the model, removing model elements corresponding to ``removed'' nodes from the model, changing attributes of ``preserved'' nodes given via the change nodes of the simple change graphs, and finally connecting the new model elements to the preserved nodes according to the edges in the simple change graph (and removing dangling edges after removing likewise). Alternatively, tools such as \textsc{Henshin} can be used to define an edit rule corresponding to the simple change graph first and then applying this edit rule as an \eo{} using the matching defined by the incomplete \eo{} $\varepsilon$.

In our implementation and evaluation, we realize the operator $\operatorname{SCG}$ as follows: For two (successive) model $m_1$ and $m_2$ (i.e., $(m_1, m_2) \in \mathcal{T}$), we compute the simple change graphs in $\mathcal{G}$ as described in \secref{sec:background}. That is, we first compute a model difference by the tool \textsc{SiDff}~\cite{schmidt2008constructing} (other model diffing tools, e.g., \textsc{EMFCompare}, are conceivable). We then map added, removed, and changed model elements to their corresponding nodes in a simple change graph. We then remove all nodes from the matching tree not directly connected to these added, removed, or changed nodes.

\begin{remark}
Only depending on the matching defined by the incomplete \eo{} brings some limitation to the approach. During the software model completion, we can not depend on ``new'' preserved nodes. Anyway, it is hardly possible, to include the entire model in the context, one therefore needs to take decisions to only bring in slices of the model into the model completions. Alternative slicing options or extensions of the simple change graphs used in this work are conceivable. For example, one could try to extend the simple change graph with preserved nodes that are likely be involved in a subsequent model completion (e.g., model elements that are textually or semantically similar to the elements involved in the current change context). Consideration of these alternatives are beyond the scope of the present work and left for future work.
\end{remark}
}{}

\subsection{Details for Baseline Comparison}

\subsubsection{Evaluated Datasets}
\changedFinalRevision{

A proper level to compare different model completion approaches would be a meta-meta level (following the MOF). This allows us to compare Chaaben et al.'s approach, which works for a subset of UML class diagrams and a subset of activity diagrams, against our approach, which works directly on the abstract syntax graph of the models. By interpreting the abstract syntax of our models as class diagrams, we were able to compare (part of) their approach to ours. 
In our dataset, we had already classified the changes and the changes that are of interest for the recommendation of new concepts, corresponded to a class of changes we called "Add\_node" in our samples. We selected these changes, which left us with 51 samples for the revision dataset.

\subsubsection{Evaluation Method}

In our comparison, we focused on concept recommendation and association recommendation. Attribute recommendation in class diagrams is quite comparable to concept recommendation. Indeed, in \textsc{Ecore} deciding between a reference to another concept or an attribute of an EClass is more like a design choice and both are considered to be a EStructuralFeature.

We mapped the examples we observed to corresponding concept recommendations. E.g., when an EClass with name User is added via a containment to an EClass with the name Software, we used the tuple [EClass.Software, EClass.User] in Chaaben et al.'s approach. Depeding on the concrete ECore concept, we replaced *name* by the corresponding identifier (e.g., name for EClass, key for EAnnotation, etc.).

To get a clearer picture of the pros and cons of the approaches, we decided to report independently the accuracy of the correct concepts being recommended, the accuracy of correct association being recommended, and we further split in correct type (e.g., EClass in the example above) and correct name (e.g., Software, or User in the example above).

\subsubsection{Replication of their approach}

To use the approach introduced by Chaaben et al.~\cite{chaaben2023towards}, as the \textsc{Baseline} we had to make some design decisions, to make a fair comparison possible.

We utilized the same few-shot examples, following the premise that these could originate from unrelated models. We could have decided to choose the few-shot samples from the dataset, similar to our approach. Anyway, since we consider this a crucial difference of the approach, we used the few-shot samples exactly as in the implementation of their approach. The few-shot samples were loaded from a file that we used in the re-implementation of their approach.
We build on their serialization of concepts and enhance the queries by additionally incorporating the partial domain model in a similar manner. We select between one to four pairs of related concepts, enclosing the concept names in brackets.


As in the original approach, we query \textsc{GPT-3} (text-davinci-002) multiple times to suggest the most frequently occurring concepts. We use the same temperature setting of 0.7 and a maximum token length of 20 tokens, which is sufficient in our \textsc{Revision} dataset to suggest at least one new pair of concepts. Excess tokens were removed. 
We also considered upgrading the model used for the \textsc{Baseline} to GPT-4. This transition would necessitate additional modifications in their approach. When utilizing GPT-4 with the existing prompts, the model begins generating natural language text that is not directly related to the specific use case. This deviation occurs because the text-davinci-002 model is not designed for chat-like interactions, unlike ChatGPT and GPT-4. Consequently, changing the model would require a redesign of their prompts to align with the capabilities of these models.

The authors employed a sampling strategy coupled with a ranking method, so we similarly query GPT-3 multiple times using a variety of prompts, each consisting of the same few-shot examples but with different queries that incorporate a subsets of model elements from the partial model. In their implementation, the authors sampled (random or all) pairs of concepts from the model at hand. 
This method does not facilitate effective real-time responses, particularly for larger software models, thus making it impractical for scenarios like those encountered in our \textsc{Revision} dataset (about 685 queries on average, see \ref{tbl:datasets}).
Since we had simple change graphs at hand, we decided to sample the edges/associations from these simple change graphs, ensuring that the number of elements in the partial model for each query ranges between one and four elements.


Furthermore, when suggesting new concepts, the paper considers both elements of each pair as new concepts. Unfortunately, one of these elements is usually already present in the partial model. This results in existing model elements being ranked at the top, rather than new concepts, leading to poorer outcomes. We have improved upon this by filtering out classes that already exist in the model.

It is noteworthy that several recommender systems make a list of k recommendations. We did intentionally decide to set k equal to one, since in a real world scenario (compare to GitHub Copilot), one would typically not come up with a list, but directly integrate one recommendation in the IDE. To ensure a fair comparison, we focus on the single most frequently occurring completion.

}{}


\subsection{Few-shot examples}\label{sec:fewsohtexamples}

Due to space limitations, we omitted the few-shot examples in Figure \ref{fig:motivatingexample} which are presented here. These examples demonstrate the retrieval of, in this specific case, four few-shot instances through our vector store. The  similarity-based retrieval mechanism is further detailed in \secref{sec:semanticretrieval}.

\begin{lstlisting}[frame=single, basicstyle=\sffamily\scriptsize, breaklines=true, breakindent=0pt]
t # 5175
e 2 1 "{'changeType': 'Add', 'type': 'reference', 'referenceTypeName': 'eOperations'}" "{'changeType': 'Preserve', 'type': 'object', 'className': 'EClass', 'attributes': {'id': '_ftfz6d6tEei97MD7GK1RmA', 'eAnnotations':['org.eclipse.emf.ecore.impl.EAnnotationImpl@1d8d14f1 (source: http://www.eclipse.org/emf/2002/GenModel)','org.eclipse.emf.ecore.impl.EAnnotationImpl@c8ca1dd (source: duplicates)'],'name':'Classifier','ePackage':'uml','abstract':'true','interface':'false', 'eIDAttribute':'name','eStructuralFeatures':['isAbstract','generalization','powertypeExtent','feature','inheritedMember','redefinedClassifier','general','substitution','attribute','representation','collaborationUse','ownedUseCase','useCase'],'eGenericSuperTypes':['org.eclipse.emf.ecore.impl.EGenericTypeImpl@239c2926 (expression: Namespace)','org.eclipse.emf.ecore.impl.EGenericTypeImpl@526bc7ba (expression: RedefinableElement)','org.eclipse.emf.ecore.impl.EGenericTypeImpl@6999e7c8 (expression: Type)','...']}}" "{'changeType': 'Add', 'type': 'object', 'className': 'EOperation', 'attributes': {'id': '_mrycqN6tEei97MD7GK1RmA', 'name':'getAllUsedInterfaces','ordered':'false','unique':'true','lowerBound':'0','upperBound':'-1','many':'true','required':'false','eType':'Interface','eGenericType':'org.eclipse.emf.ecore.impl.EGenericTypeImpl@762545f6 (expression: Interface)','eContainingClass':'Classifier'}}"
e 2 0 "{'changeType': 'Add', 'type': 'reference', 'referenceTypeName': 'eOperations'}" _ "{'changeType': 'Add', 'type': 'object', 'className': 'EOperation', 'attributes': {'id': '_mrycp96tEei97MD7GK1RmA', 'name':'getUsedInterfaces','ordered':'false','unique':'true','lowerBound':'0','upperBound':'-1','many':'true','required':'false','eType':'Interface','eGenericType':'org.eclipse.emf.ecore.impl.EGenericTypeImpl@3d23f56e (expression: Interface)','eContainingClass':'Classifier'}}"

$$
---
t # 1250
e 2 1 "{'changeType': 'Add', 'type': 'reference', 'referenceTypeName': 'eOperations'}" "{'changeType': 'Preserve', 'type': 'object', 'className': 'EClass', 'attributes': {'id': '_ftfz6d6tEei97MD7GK1RmA', 'eAnnotations':['org.eclipse.emf.ecore.impl.EAnnotationImpl@50bd114f (source: http://www.eclipse.org/emf/2002/GenModel)','org.eclipse.emf.ecore.impl.EAnnotationImpl@11c9b440 (source: duplicates)'],'name':'Classifier','ePackage':'uml','abstract':'true','interface':'false','eIDAttribute':'name','eStructuralFeatures':['isAbstract','generalization','powertypeExtent','feature','inheritedMember','redefinedClassifier','general','ownedUseCase','useCase','substitution','attribute','representation','collaborationUse','ownedSignature'],'eGenericSuperTypes':['org.eclipse.emf.ecore.impl.EGenericTypeImpl@1504a6f7 (expression: Namespace)','org.eclipse.emf.ecore.impl.EGenericTypeImpl@65db7f4d (expression: RedefinableElement)','org.eclipse.emf.ecore.impl.EGenericTypeImpl@225a383c (expression: Type)','...']}}" "{'changeType': 'Add', 'type': 'object', 'className': 'EOperation', 'attributes': {'id': '_inuJYt6tEei97MD7GK1RmA', 'name':'getOperation','ordered':'false','unique':'true','lowerBound':'0','upperBound':'1','many':'false','required':'false','eType':'Operation','eGenericType':'org.eclipse.emf.ecore.impl.EGenericTypeImpl@4e5c0171 (expression: Operation)','eContainingClass':'Classifier','eParameters':['name']}}"
e 1 0 "{'changeType': 'Add', 'type': 'reference', 'referenceTypeName': 'eParameters'}" _ "{'changeType': 'Add', 'type': 'object', 'className': 'EParameter', 'attributes': {'id': '_inuJY96tEei97MD7GK1RmA', 'name':'name','ordered':'false','unique':'true','lowerBound':'1','upperBound':'1','many':'false','required':'true','eType':'String','eGenericType':'org.eclipse.emf.ecore.impl.EGenericTypeImpl@bbcf831 (expression: String)','eOperation':'getOperation'}}"

$$
---
t # 2292
e 0 2 "{'changeType': 'Remove', 'type': 'reference', 'referenceTypeName': 'eAnnotations'}" "{'changeType': 'Preserve', 'type': 'object', 'className': 'EClass', 'attributes': {'id': '_fthA796tEei97MD7GK1RmA', 'eAnnotations':['org.eclipse.emf.ecore.impl.EAnnotationImpl@2fa33653 (source: http://www.eclipse.org/emf/2002/GenModel)','org.eclipse.emf.ecore.impl.EAnnotationImpl@59d423ca (source: duplicates)'],'name':'Extension','ePackage':'uml','abstract':'false','interface':'false','eOperations':['non_owned_end','is_binary','getStereotype','getStereotypeEnd','isRequired','getMetaclass','metaclassEnd'],'eStructuralFeatures':['isRequired','metaclass'],'eGenericSuperTypes':['org.eclipse.emf.ecore.impl.EGenericTypeImpl@3ff99636 (expression: Association)']}}" "{'changeType': 'Remove', 'type': 'object', 'className': 'EAnnotation', 'attributes': {'id': '_oBpkOd6tEei97MD7GK1RmA', 'source':'http://www.eclipse.org/emf/2002/GenModel','details':['org.eclipse.emf.ecore.impl.EStringToStringMapEntryImpl@95f12e0 (key: documentation, value: An extension is used to indicate that the properties of a metaclass are extended through a stereotype, and gives the ability to flexibly add (and later remove) stereotypes to classes.)'],'eModelElement':'Extension'}}"
e 2 3 "{'changeType': 'Remove', 'type': 'reference', 'referenceTypeName': 'details'}" _ "{'changeType': 'Remove', 'type': 'object', 'className': 'EStringToStringMapEntry', 'attributes': {'id': '_oBpkOt6tEei97MD7GK1RmA', 'key':'documentation','value':'An extension is used to indicate that the properties of a metaclass are extended through a stereotype, and gives the ability to flexibly add (and later remove) stereotypes to classes.'}}"
e 0 4 "{'changeType': 'Add', 'type': 'reference', 'referenceTypeName': 'eAnnotations'}" _ "{'changeType': 'Add', 'type': 'object', 'className': 'EAnnotation', 'attributes': {'id': '_0oByC96tEei97MD7GK1RmA', 'source':'http://www.eclipse.org/emf/2002/GenModel','details':['org.eclipse.emf.ecore.impl.EStringToStringMapEntryImpl@5cd02377 (key: documentation, value: An extension is used to indicate that the properties of a metaclass are extended through a stereotype, and gives the ability to flexibly add (and later remove) stereotypes to classes.\\n<p>Merged from package UML (URI {@literal http://www.omg.org/spec/UML/20110701}).</p>)'],'eModelElement':'Extension'}}"
e 4 1 "{'changeType': 'Add', 'type': 'reference', 'referenceTypeName': 'details'}" _ "{'changeType': 'Add', 'type': 'object', 'className': 'EStringToStringMapEntry', 'attributes': {'id': '_0oByDN6tEei97MD7GK1RmA', 'key':'documentation','value':'An extension is used to indicate that the properties of a metaclass are extended through a stereotype, and gives the ability to flexibly add (and later remove) stereotypes to classes.\\n<p>Merged from p'}}"

$$
---
t # 88
e 0 2 "{'changeType': 'Add', 'type': 'reference', 'referenceTypeName': 'eAnnotations'}" "{'changeType': 'Preserve', 'type': 'object', 'className': 'EClass', 'attributes': {'id': '_fZD13N6tEei97MD7GK1RmA', 'eAnnotations':['org.eclipse.emf.ecore.impl.EAnnotationImpl@68481491 (source: http://www.eclipse.org/emf/2002/GenModel)','org.eclipse.emf.ecore.impl.EAnnotationImpl@4faef368 (source: duplicates)'],'name':'DataType','ePackage':'cmof','abstract':'false','interface':'false','eIDAttribute':'name','eStructuralFeatures':['ownedOperation','ownedAttribute'],'eGenericSuperTypes':['org.eclipse.emf.ecore.impl.EGenericTypeImpl@7ebe7d9f (expression: Classifier)']}}" "{'changeType': 'Add', 'type': 'object', 'className': 'EAnnotation', 'attributes': {'id': '_ffDLSt6tEei97MD7GK1RmA', 'source':'http://www.eclipse.org/emf/2002/GenModel','details':['org.eclipse.emf.ecore.impl.EStringToStringMapEntryImpl@5e553e0a (key: documentation, value: A data type is a type whose instances are identified only by their value. A data type may contain attributes to support the modeling of structured data types.)'],'eModelElement':'DataType'}}"
e 2 1 "{'changeType': 'Add', 'type': 'reference', 'referenceTypeName': 'details'}" _ "{'changeType': 'Add', 'type': 'object', 'className': 'EStringToStringMapEntry', 'attributes': {'id': '_ffDLS96tEei97MD7GK1RmA', 'key':'documentation','value':'A data type is a type whose instances are identified only by their value. A data type may contain attributes to support the modeling of structured data types.'}}"
e 3 4 "{'changeType': 'Remove', 'type': 'reference', 'referenceTypeName': 'details'}" "{'changeType': 'Remove', 'type': 'object', 'className': 'EAnnotation', 'attributes': {'id': '_fZD13d6tEei97MD7GK1RmA', 'source':'http://www.eclipse.org/emf/2002/GenModel','details':['org.eclipse.emf.ecore.impl.EStringToStringMapEntryImpl@77d8e24f (key: documentation, value: A data type is a type whose instances are identified only by their value. A DataType may contain attributes to support the modeling of structured data types.\\n\\n\\n\\nA typical use of data types would be to represent programming language primitive types or CORBA basic types. For example, integer and string types are often treated as data types.\\r\\nDataType is an abstract class that acts as a common superclass for different kinds of data types.)'],'eModelElement':'DataType'}}" "{'changeType': 'Remove', 'type': 'object', 'className': 'EStringToStringMapEntry', 'attributes': {'id': '_fZD13t6tEei97MD7GK1RmA', 'key':'documentation','value':'A data type is a type whose instances are identified only by their value. A DataType may contain attributes to support the modeling of structured data types.\\n\\n\\n\\nA typical use of data types would be to'}}"
e 0 3 "{'changeType': 'Remove', 'type': 'reference', 'referenceTypeName': 'eAnnotations'}" _ _

$$
---


\end{lstlisting}

\subsection{Further (pre-)processing and filtering steps}
We perform some additional filtering steps during the \mbox{(pre-)}processing of simple change graphs and the sampling. For the sake of clarity, we omitted them in the description of the approach and experiment description. The applied filters are the following:

\begin{itemize}
    \item Because we have a limitted context size available for the LLMs, very long attribute descriptions (for example in comments) are limitted to a length of 200 characters. Everything longer than 200 characters has been cut and ``\dots'' are appended.
    \item When sampling few-shot samples, and the overall prompt size becomes too long, we remove few-shot samples until the prompt fits into the model.
    \item Serialized \scg{s} that are too large to fit in the context of the language model are filtered.
    \item To save tokens and therefore reduce language model usage costs, we do not repeat node labels, but instead replace them by a ``\_'' token (or ``\{\}'' when using JSON in the label representation).
    \item We filtered duplicated \scg{s}.
    \item Models from the original \textsc{RepairVision} dataset that could not be loaded or had empty history were removed. The description of the dataset parameters in \secref{sec:dataset} describes the state after this filtering.
\end{itemize}

\changedFinalRevision{
\subsection{Detailed Results of the Industry Dataset and Experiment 3}
Table \ref{tbl:results-exp3-long} summarizes the results of our results from Experiment 3. We distinguished between four completion task characteristics in the rows of the table: noise present, project specific change, complex change, and reoccurring pattern. All four are binary relations.
``Noise Present'' indicates whether there are changes entangled in the task or in the few-shot examples. We considered the task to be a ``Project Specific Change'', if the change was not common for the modeling language (SysML), but rather some pattern we observed for this project only.
``Complex Change'' indicates a change that does consist of several interconnected atomic changes (e.g., adding an attribute or adding a class, i.e., implicitly we distinguish between atomic changes and complex changes, as common in the field~\cite{Kehrer2016}). Finally ``Reoccurring Pattern'' describes if we observe the pattern of the task at hand also in the few-shot samples. That is, aside from concrete attribute values, the change happens often in the project and can be retrieved via our semantic retrieval. Correctness is classified as follows in the columns of the table: We only consider semantically correct completions (evaluated via a manual analysis) as \emph{correct}. The \emph{incorrect} completions, we further classify in ``semantically conceivable'' (i.e., the change is not the one observed in the ground truth, but without further context it would also be meaningful), ``Semantically Incorrect'' (i.e., format correct, structurally correct, but the completion has a meaning different from the ground truth completion), ``structurally incorrect'' (i.e., a reference connects no the right nodes, or a new class is associated to another class where nothing should be added, etc.), and ``format incorrect'' (i.e., without error correction, the graph serialization could not be parsed, e.g., because an existing node id is reused by another node).
}{}
\begin{table*}[ht]
\centering
\caption{Comparison of different failure types along several characteristics of the completion task. This table summarizes the results of our manual analysis of the \textsc{Industry} dataset.}
\label{tbl:results-exp3-long}
\begin{tabular}{@{}lcccccccc@{}}
\toprule
 & & \multicolumn{5}{c}{Level of Correctness} & \\ 
 \cmidrule(lr){3-7}
 Task Characteristic& & Correct & \multicolumn{4}{c}{Incorrect} & Total & Total (\%)\\ 
 \cmidrule(lr){3-7}
   &  & Semantically  & Semantically  & Semantically  & Structurally  & Format  &  & \\ 
 &  & Correct  & Conceivable   & Incorrect     & Incorrect     & Incorrect    &  & \\
\midrule
\textbf{Noise Present} & TRUE & 30.77\% & 23.08\% & 15.38\% & 15.38\% & 15.38\% & 13 & 11\\

                       & FALSE & 66.06\% & 14.68\% & \phantom{1}8.26\% & \phantom{1}4.59\% & \phantom{1}6.42\% & 109 & 89\\
\midrule
\textbf{Project Specific Change} & TRUE & 74.55\% & \phantom{1}3.64\% & \phantom{1}7.27\% & \phantom{1}5.45\% & \phantom{1}9.09\% & 55 & 45\\

                                      & FALSE & 52.24\% & 25.37\% & 10.45\% & \phantom{1}5.97\% & \phantom{1}5.97\% & 67 & 55\\
\midrule
\textbf{Complex Change} & TRUE & 66.67\% & 23.81\% & \phantom{1}4.76\% & \phantom{1}0.00\% & \phantom{1}4.76\% & 21 & 17\\
                             & FALSE & 61.39\% & 13.86\% & \phantom{1}9.90\% & \phantom{1}6.93\% & \phantom{1}7.92\% & 101 & 83\\
\midrule
\textbf{Reoccurring Pattern} & TRUE & 71.13\% & 17.53\% & \phantom{1}6.19\% & \phantom{1}2.06\% & \phantom{1}3.09\% & 97 & 80\\
                             & FALSE & 28.00\% & \phantom{1}8.00\% & 20.00\% & 20.00\% & 24.00\% & 25 & 20 \\
\midrule
\textbf{Total} & & 62.30\% & 15.57\% & \phantom{1}9.02\% & \phantom{1}5.74\% & \phantom{1}7.38\% & 122 & 100\\
\bottomrule
\end{tabular}
\end{table*}

\changedFinalRevision{
\begin{remark}
In \secref{sec:evaluation}, we stated that there is no significant relationship between the correctness and the number of few-shot examples that are added to the prompt.
For the \textsc{Industry} dataset, we additionally recorded, if (at least one) similar pattern is among the few-shot examples. Separating these two cases -- i.e., there is a similar pattern among the few-shot examples or not -- we see that in the first case there is no significant relationship between the number of few-shot examples, while in the second case there is a significant relationship. 

This suggest that as long as a similar few-shot example is available, the amount of few-shot examples does not matter too much, while in the case that the examples are rather unrelated, the amount plays a role.
\end{remark}

}{}

\subsection{Detailed Results of the Fine-Tuning Experiments}\label{sec:detailsexperiment}

This section delves into a detailed analysis of our last experiment highlighting the influence of various factors on the \emph{average token accuracy}.
We are especially interested in the model token accuracy of the fine-tuned language model in relationship with the properties of the dataset and the properties of the fine-tuning such as the number of fine-tuning epochs and the base language model used. 
We fine-tune one LLM per simulated repository. As base models we choose \ada{}, \curie{}, and \davinci{} from the GPT-3 family. 
Since fine-tuning the \davinci{} model is quite expensive (i.e., 3 Cents per thousand tokens at the time of this experiment), we fine-tuned this model only for the model repositories where the perturbation probability equals 100\% (the ones which are typically the harder ones). This leaves us with a total of 112 fine-tuned models (24 simulated repositories for \ada{} and \curie{} and 8 for \davinci{}, which is 24*2*2+ 8*2*1 = 112) and a total fine-tuning cost of \totalCost{}. 
Building on the insights previously touched upon, our analysis reveals a strong correlation between average token accuracy and the number of fine-tuning epochs. Furthermore, it becomes evident that larger models exhibit better performance in terms of average token accuracy. 
Regarding the repository properties, we only find significant negative correlations with the perturbation probability (Table \ref{tbl:exp3_corr}).
We therefore also analyze model completions from a graph matching perspective (like already mentioned in Experiment 4). Since generating all completion candidates for all test samples of all fined-tuned language models would be even more expensive, we select two fine-tuned language models, the less cost-intensive alternative, and perform the analysis of the model completions on them. 

\begin{table}[htbp] 
\caption{Pearson correlations of the average token accuracy w.r.t. several properties. Repo$_D$ denotes the number of revisions, Repo$_E$ the number of applied \eo{s}, and Repo$_P$ the perturbation probability.}
\begin{adjustbox}{center}
    \scriptsize
    \begin{tabular}{r|c c c c c c}
    \toprule
     & Repo$_D$ &  Repo$_E$	& Repo$_P$ & Epochs & Token Count & Base Model \\
     \midrule
     Average Token Accuracy & \cellcolor[gray]{0.6}0.16 & \cellcolor[gray]{0.6}0.13 & \cellcolor[gray]{0.8}-0.22* & \cellcolor[gray]{0.4}0.69** &\cellcolor[gray]{0.65}0.08 & \cellcolor[gray]{0.5}0.43**\\
    Token Accuracy (All) & 0.16 & 0.13 & -0.22* & 0.69** & 0.08 & 0.43**\\
    Token Accuracy (Ada) & 0.26 & 0.22 & -0.43* &	0.72** & 0.14 & -- \\
     Token Accuracy (Curie) & 0.13 & 0.12 & -0.35* & 	0.82** & 0.02 & -- \\
     Token Accuracy (Davinci) &	0.02 & -0.04 & -- &	0.94** &	-0.06 &	-- \\
    \bottomrule
    \end{tabular}

  \end{adjustbox}
        \begin{flushright}
      \footnotesize
       (**: $p < .001$, *: $p<0.05$)
       \end{flushright}
\label{tbl:exp3_corr}

\end{table}
\end{appendix}

\subsection{Related Work }

\begin{sidewaystable*}


\caption{Related work summary.}

\begin{adjustbox}{max width=\textwidth}
\footnotesize
\begin{tabular}
{p{0.7cm}p{3.3cm}p{3.0cm}p{5cm}p{3cm}p{4cm}p{1cm}p{5cm}}
\toprule
Paper & Task  & Method & Evaluation & Data & Prerequisites (for evaluation) & History & Comparison Possible?\\
\midrule
\cite{adhikari2023simima} \cite{stephan2019towards} & Single-step operations and similar, related Simulink systems & Information retrieval (association rule mining, frequency-based matching) & Metrics analysis (prediction, accuracy, error classification) & Simulink (available) & None & No & With adaption: Adaption of \rag{} to Simulink datasets beyond the scope.\\
\hline
\cite{heinemann2012facilitating} & Library block recommendation & Information retrieval (association rules, collaborative filtering) & Metrics analysis (precision, recall, and F-measure) & Simulink (available) & None & No  & With adaption: Adaption of \rag{} to Simulink datasets beyond the scope.\\
\hline
\cite{agt2018domore} \cite{agt2019automated} & Recommendation of related classes, possible sub- or super-classes, relationships between elements, element names & Knowledge graphs, semantic web technologies & Planned user study but not yet conducted (no metric) & UML (not available) & Conceptual knowledge bases (semantically related terms, built from natural language data) and semantic network (not available) & No & No: Dataset and (parts of) their approach are not available.\\
\hline
\cite{deng2016recommendation} & Activity node recommendation & Information retrieval (similarity-based, pattern mining, pattern as relationships between activity nodes) & Metrics analysis (HitRate, Precision, Recall, and F1 Score) & Business process modeling (not available) & Database constructed from existing processes (not available) & No & No: Their dataset is not available to us and the approach is domain specific (BPMN).\\
\hline
\cite{di2023memorec} & Recommendation of entities in metamodels (classes, structural features), no support for types of the recommended attributes, relationships & Information retrieval (similarity-based, collaborative filtering strategy) & Metrics analysis (rather best case scenario -- out of N items some of the (possibly huge) model are correct -- (success rate (SR@N), precision, recall, F1 score)) & Ecore metamodels (available) & Predefined categories/labels beneficial & No & With adaption: 
Adaption of their approach to our history-based data possible but beyond the scope of this work. Note: Only sub-tasks of model completion.\\
\hline
\cite{elkamel2016uml} & Recommendation of new concepts (i.e., class names), attributes, operations & Clustering algorithm (on semantic relations) & User study (relevant and new recommendations (PN), non useful recommendations (NU) not recommended but included in individual design (NR), relevant rate (TP), accuracy rate of new suggestions (TN)) & UML (not available) & Clustered UML diagrams (not available) & No & No: Use cases are slightly different. They recommend classes and adapt with user feedback, we recommend model elements (in ``small steps''). \\
\hline
\cite{kogel2016automatic} & Model completion & Pattern matching and Association rule mining & Metrics analysis (Precision) & Eclipse GMF Project meta-models (available) & Catalog of change patterns (not available) & Yes & No: We could apply our approach to their data, but their numbers are reported based on the concept of edit rule applications and therefore can not be compared. Anyway, an adaption and reimplementation of their evaluation seems reasonable but is beyond the scope.\\
\hline
\cite{kuschke2017rapmod} \cite{kuschke2013recommending} \cite{maeder2021} & Model completion & Rule-based pattern matching & User study (number of saved user actions, time against manual completion) & UML (not available) & Catalog of change patterns (not available)& Yes & No: Dataset used for evaluation is not available to us and running their approach on our data requires a catalog of change patterns that is not available. As for~\cite{kogel2016automatic}, reimplementation of evaluation seems reasonable but is beyond the scope. \\
\hline
\cite{di2023morgan}  & Recommending new concepts (class names) and attributes & Information retrieval (similarity-based, graph kernels, TF-IDF) & Metrics (success rate, precision, recall, and F-measure (modified)) analysis (best-case scenario, i.e., check whether one out of N recommendations correct -- evaluated on ``token'' level, structural correctness, e.g., a new class connected to $>=2$ other classes, not evaluated and not reflected in the approach) & ModelSet, reverse engineered class diagrams from Java code, JSON crawled from GitHub, Ecore metamodels (partially available) & None & No &  With adaption: Structural correctness not reflected by their approach. Adapting their approach to compare for concept and attribute recommendation seems reasonable. Note: Only sub-tasks of model completion.\\
\hline
\cite{chaaben2023towards} & Recommending new concepts (class names), attributes, association names & Machine Learning (GPT-3, few-shot learning) & Metrics analysis (30 models selected and evaluated manually, precision, recall) & ModelSet (available) & None & No & With adaption: We adapted their approach to work on EMF-based models (interpreting them as class diagrams).\\
\hline
\cite{burgueno2021nlp} &  ``Contextualized'' model completion & Machine Learning (reuse pre-trained word embedding models, project-specific training, NLP-based system, word embedding similarity based on textual information) & Metric analysis (Precision, Recall) -- best case scenario, out of N items & Industrial data (incident management system in municipal water supply and sewage in Malaga)(not available) & Textual information required (of project and/or related business domain)(not available) & No & No: Their dataset not available to us and other artifacts not (explicitly) available in our dataset.\\
\hline
\cite{di2022finding}  & Recommendation of edit operations given preceding edit operations (on a type level, i.e., omitting attribute values) & Machine Learning (Encoder-Decoder LSTM neural network) & Metric analysis, limited possibilities due to ignoring names and values, out of N items some are correct, unclear how many items get recommended (Success rate, precision recall) & BPMN  & None & Yes & No: Their approach depends on operation recording, which we do not have available for our datasets. Regarding their BPMN dataset, we could not find models (+ metamodel) for an application of our approach to their dataset. \\


\bottomrule
\end{tabular}
\end{adjustbox}

\label{tbl:relatedwork}
\end{sidewaystable*}





\subsubsection{Current Challenges in the Research Domain}

Research in \mde{} faces several challenges that should receive increased attention in the future.

The scarcity of reusable datasets~\cite{damasceno2021quality,robles2023reflection,lopez2022modelset,bucchiarone2020grandchallenges} for many use cases in \mde{} hinders the comparison of different approaches, 
which is then often reduced to a qualitative analysis.
The lack of proper datasets also poses a challenge for the development and evaluation of data-driven (e.g., machine learning) approaches in \mde{}.
To circumvent this lack of datasets, many authors in \mde{} research report on experiences using their approaches in a concrete application context, that is, as part of a tool. 
Reporting on an evaluation in a concrete application setting, again, makes it difficult to compare against the approach, especially if the application context or the tool is not available to the public and/or user studies are performed.

There are only a few datasets available that can be used to evaluate model completion.
In the concrete example of model completion, the evaluation is often performed on a dataset of model snapshots, from which elements are removed artificially. 
Instead, it would be more realistic to have pairs of to-be-completed models and their completed counterparts.

Finally, there are no commonly accepted evaluation metrics and often technologies or proposed approaches are evaluated in a manner that is only applicable for the specific use case at hand. Only a minority of the literature reports on metrics that are independent of their specific approach and only depending on the use case (i.e., model completion). 
For instance, a model completion methodology could suggest the top-10 names for meta-model classes for inclusion in a meta-model, with the evaluation of this method focusing solely on the accuracy of these ten recommendations. Consequently, this creates a challenge in directly comparing such an approach to others that might recommend a single name while also suggesting relationships between the newly added class and existing classes.
Further it would require a ground truth of to-be-completed and completed models, which is not available for most datasets.
But even here, its not easy to define what a correct completion is. For example, if a model element is missing in the incomplete model, but the model element is not required for the model to be valid, is it a correct completion or not?
Likewise, if a recommended class name is a synonym of the correct class name, is it a correct completion or not?
Note that for source code, there are commonly accepted datasets such as HumanEval~\cite{chen2021codex} and evaluation metrics~\cite{chen2021codex} to evaluate code completion approaches.
For example, since there is a well-defined execution semantics, the evaluation of a code completion approach can be performed by checking the correctness of the code completion in a test suite.
For many models (e.g., UML, SysML, Ecore, etc.), there is no well-defined execution semantics and therefore a test approach for evaluation would not be applicable to software models, in general. 
\changedFinalRevision{
\subsubsection{Comparison and differentiation from other approaches}

In Table \ref{tbl:relatedwork}, we summarize the related work with a specific focus on the model completion task. 
For each approach, we included information about the specific task, the method used and the evaluation process, including the data used for evaluation and specific prerequisites are required for replicating the evaluation. Given the sometimes challenging nature of tracking the availability of artifacts, we acknowledge that some information might not be entirely accurate, and we apologize for any inadvertent inaccuracies. 
A main finding of our analysis of related work is that, currently, a direct comparison with other approaches, for most approaches, is infeasible, due to the field's novelty, the absence of commonly accepted metrics and datasets, or a focus of previous work on artificial datasets without real-world model evolution. We also highlight the most prevailing reason why a comparison to our approach is infeasible in Table \ref{tbl:relatedwork}. In the following paragraph, we will delve into the reasons why a direct comparison to the approaches in Table \ref{tbl:relatedwork} is not easily possible. 

When we want to compare our approach to an existing approach, we could do this comparison on he data the approach was evaluated on. For this, the exact (test) data and a comparable metric need to be published. If this is not the case, it might still be possible to perform a comparison by applying the existing approach to our data. But for this the approach should either be generic (i.e., not depend on a specific domain) or work for our datasets (i.e., EMF models).

\paragraph{(Partial) model completion}


The approach by Agt-Rickauer et al. \cite{agt2018domore,agt2019automated} focuses on suggesting related class names, potential sub- or super-class names, and different names for connections given a specific focus point in the model. However, their approach does not extend to suggesting attributes, operation names, or relationship types, making a comparison to our approach challenging. Conversely, our approach goes far beyond suggesting the names of new elements. Furthermore, the evaluation of their method is deeply integrated within the tool, making a direct comparison impossible. This approach relies on conceptual knowledge bases (comprising semantically related terms built from natural language data) and a semantic network, neither of which are available for external validation. Consequently, it is challenging to verify when and how the suggestions are semantically and structurally correct.

The approach by Elkamel et al. \cite{elkamel2016uml} suggests entities in metamodels, such as classes and structural features, but does not support types for the recommended attributes or relationships. In an offline phase, they use a clustering algorithm to partition UML classes collected from various UML class diagrams based on the semantic relations between their characteristics. Subsequently, they recommend semantically similar whole classes, and individual methods and attributes of that class can be accepted or rejected.
Their approach is not based on historical data. On the other hand, we cannot apply their approach to our data, as they only suggest entire classes while our approach focuses on a more general setting. 
Their approach directly relies on user feedback, as entire classes are suggested for the user including its attributes to accept or reject. As a result, it is rare for an entire class to be completely correct initially. While their focus is more on the user setting, we focus on the core effectiveness of the LLM technology.
This makes a direct comparison between our methods unintuitive.


Similarly, Di Rocco et al. \cite{di2023memorec} propose an approach for suggesting new classes and structural features (attributes and references) in metamodels. Their method generates recommendations as a ranked list of classes if the active context is a package, or as a ranked list of structural features if the active context is a class. This approach involves identifying a subset of the most similar metamodels from given metamodel repositories and determining the most similar contexts within that subset. However, the method lacks support for recommending the types of attributes and relationships.



Di Rocco et al.~\cite{di2022finding} further present a recommender system that uses an Encoder-Decoder neural network to assist modelers with performing editing operations. The system learns from past modeling activities and is evaluated on a BPMN dataset. 
These past activities are modeled as edit operation sequences. One limitation of this specific format is that the changes of an element in the edit operation sequences can be scattered throughout the complete sequence, with possibly hundreds of other edit operations between them. This also means that connected/related elements or elements that belong together can appear at completely different locations within such a sequence. These connected/related elements can give valuable context to the model completion task. If then, as in the work by Di Rocco et al.~\cite{di2022finding}, only the last 10 edit operations are considered, important information regarding the local (graph-like context) might be lost. 
This issue becomes more pronounced as models increase in size. One could instead not only focus on the last x edit operations, but instead put the entire history of a model into the LLM context. Anyway, especially for large models, providing the entire history of the model as context is infeasible and may not fit in the context of an LLM. 

We cannot compare their approach to our model completion approach because it does not include the specific details and values of operations. For instance in the example in Listing \ref{list:nemo}. 

}{}
\begin{lstlisting}[frame=single, caption={An example of the NEMO \cite{di2022finding}}, basicstyle=\sffamily\scriptsize, breakindent=1pt, captionpos=b, label={list:nemo}, breaklines=true, postbreak=\mbox{\textcolor{red}{$\hookrightarrow$}\space}]
set-att name BPMN2ActionContributor to #200
\end{lstlisting}
\changedFinalRevision{
In the \lstinline[basicstyle=\sffamily\small]{setAtt}
 operation, the class name being created isn't suggested. Instead, each event is simplified to a tuple \lstinline[basicstyle=\sffamily\small]{<setAtt, class, name>}.
While their approach focuses on proposing simplified completions, as highlighted in their work, our approach suggests more complex model elements with detailed, specific values (e.g. class names, concrete attribute values). An example of a concrete, linearized model completion suggestion of \rag{} is given Listening \ref{list:Ramc}.

}{}

\begin{lstlisting}[frame=single, caption={A RAMC completion candidate}, basicstyle=\sffamily\scriptsize, breakindent=0pt, captionpos=b, label={list:Ramc}, breaklines=true, postbreak=\mbox{\textcolor{red}{$\hookrightarrow$}\space}]
1 2 "{'changeType': 'Add', 'type': 'reference',
'referenceTypeName':'eOperations'}" _ "{'changeType':'Add',
'type': 'object', 'className': 'EOperation', 'attributes'
{'id':'_lU7gFt6tEei97MD7GK1RmA','name':
'getMetaclass','ordered':'false','unique':'true',
'lowerBound':'0','upperBound':'1','many':'false',
'required':'false','eType':'Metaclass',
'eGenericType':'Metaclass','eContainingClass':
'Extension'}}"
\end{lstlisting}

\changedFinalRevision{

The competition candidate includes a specific change to the model, detailing how it is connected to other elements (as a quick reminder, 1 and 2 represent the source and target nodes, followed by the edge attributes in the first {}). It suggests specific values and the names of changed attribute elements and much more. All in all, comparing our work to the work by Di Rocco et al.~\cite{di2022finding} would be an unfair comparison for both sides.


Di Rocco et al.~\cite{di2023morgan} focus on model completion by suggesting new classes and structural features (attributes, references, methods, and fields). This is achieved by constructing a separate graph for each class in the models and use graph kernel similarity to identify the most similar items among the training set to the partial model that should be completed.
However, their approach does not ensure structural correctness, such as where to add a class or how elements should be connected overall. Additionally, the work reports very low precision and recall values, which left us believing that their method will likely not perform well on real-world and industrial datasets. For our baseline, we therefore decided to reimplement the approach by Chaaben et al.~\cite{chaaben2023towards}, which is also more closely related to our approach.

Regarding the use of language models, Chaaben et al. \cite{chaaben2023towards} use LLMs and acknowledge that their results are preliminary, considering only a few UML examples (30 domain models, selected manually from the dataset ModelSet). Their evaluation focuses on suggesting class names, attributes of classes and association names. They do not consider historical data, therefore without major adaptation, we cannot perform our experiments on their data. We primarily focus on historical data because we aim to gather real-world examples. Instead of randomly excluding model components and using them as ground truth, we study actual real-world evolution, making the setting much more realistic. Additionally, their approach does not scale for larger models, so our real-world models are by far too large to fit within the context window of the GPT-3 model (text-davinci-002). This challenge is precisely why we decided to focus on historical data in combination with simple change graph slicing. Anyway, since the work by Chaaben et al. \cite{chaaben2023towards} is the most closely rated work, we re-implement their approach and tailor it to our dataset to enable a direct comparison between their method and ours.

 }{}
\changedFinalRevision{
\paragraph{Additional data required for model completion }

There are also other approaches, such as those employing rule-based matching based on a predefined catalog of change patterns (edit operations), where additional data is required to compare their approach to ours~\cite{kuschke2017rapmod, kuschke2013recommending, maeder2021, kogel2016automatic}. These pattern-based approaches are to some extend orthogonal to ours. For example, one can use semantic lifting~\cite{Kehrer2016}, to further compress change graphs before applying an approach like ours. Similarly, one could apply an approach like ours to sequences of \eo{s}. Anyway, this requires the definition of pattern catalogs and is therefore limited to application domains where such catalogs are available or requires the combination with pattern mining~\cite{Tinnes2021,tinnes2023mining} -- an area that itself is rather active research than mature technology.

The work by Burgueno et al.~\cite{burgueno2021nlp} relies on knowledge extracted from textual documents to provide meaningful suggestions. Our approach is pure model completion and we do not include further information in the model completion setting. In this sense their approach can be understood as an extension of our approach that provides a completion in the form $C \colon \mathcal{T} \times \Sigma^{*} \to \mathcal{T}$. On the other hand, one can include other artifacts (e.g., requirements) and natural language information also as part of the model -- which is done in the form of requirements diagrams for the \textsc{SysML} models of our \textsc{Industry} dataset. In this sense -- when the additional information (e.g., requirements) that one wants to include in the model completion is fused with the to-be-completed models -- our approach covers this ``contextualized'' model completion. We acknowledge the approach by Burgueno et al.~\cite{burgueno2021nlp} and also believe that a more explicit handling of different types of context has the potential to provide better targeted model completions. Using retrieval augmented generation, further context can easily be integrated in the generation. Unfortunately, we can not easily compare their approach to ours, because the dataset used in their evaluation is not available and in our datasets we do not have this additional context readily available, rendering a direct comparison difficult.

Both research streams (i.e., contextualized model completion and pattern-based model completion) seem to be promising concepts for further investigation and extension of the approach proposed in this work. Anyway, the more ``combination'' is used in the approach the more difficult will be to understand what effect is due to which design decision or technology (without conducting a sophisticated ablation study). The purpose of this work was to investigate the merits of LLM technology for model completion and therefore increase internal validity. Further combinations are intentionally left for future research.  

\paragraph{Related but distinct tasks}

While several related studies focus on related but different tasks, we will highlight a few examples here. Of course, mentioning all of them would exceed the scope of this work, and we hope the examples given here will clarify why some related work can not be directly compared to our approach, although similar on a high abstraction level.
The work by Ohrndorf et al.~\cite{ohrndorf2021history}, which specifically proposes a model repair approach rather than model completion. Their method generates repair proposals for inconsistencies introduced by incomplete editing processes. The approach focuses on examples in the revision history, in which constraints were, at some point, violated and subsequently fixed at a later point. While their method ensures constraints are preserved, it does not handle adding or changing functionality within the system, leaving the modeler to perform the actual modeling work. As we should not compare fixing syntactic errors in source to suggesting new source code, we can also not compare model repair to model completion.
Gomes et al.~\cite{gomes2023dome} focus on creating and evolving a system domain model based on interactions in natural language from non-technical users. They utilize Natural Language Processing (NLP) to interpret the users' intents expressed in natural language and transfer these intents to commands the system can understand. This represents an entirely different task. They do not suggest functional changes to the model, but translate the user intentions to machine readable commands.
Also the approaches by Kögel et al.~\cite{kogel2016automatic} and Kushke et al.~\cite{kuschke2013recommending, kuschke2017rapmod} can be seen as different (altough very related) approaches. In principle, these approaches try to recommend complete patterns from partial patterns, which is then leveraged to perform a model completion.
Finally, the approach by Burgueno et al.~\cite{burgueno2021nlp} recommends model completions given some additional context, while our approach requires only the model (change).

\paragraph{Domain-specific applications}

There is a category of approaches focusing on specific domain languages \cite{heinemann2012facilitating, adhikari2023simima, stephan2019towards}. Their approaches are limited to Simulink models, which is why we cannot apply them to our data. 

Deng et al. \cite{deng2016recommendation} propose an approach focused on business process models, specifically BPMN. Their method is not transferable to other domains. They mine relationships among activity nodes from existing processes, store these relations as patterns in a database, and then compare new processes with these patterns. This comparison recommends suitable activity nodes from the most matching patterns to assist in building a new process.


\paragraph{The need for a benchmarking infrastructure}
Our analysis above underscores the current challenges faced by the research community. This lack of baselines and benchmarking infrastructure is a critical point. The field is currently in a state of developing the necessary infrastructure, and we are contributing to this effort, while appreciating previous contribution efforts. 

}{}

\end{document}